\documentclass[iop]{emulateapj-rtx4} 
\usepackage{natbib,amsmath,lscape}

\usepackage{lscape}
\usepackage{graphicx}
\usepackage{natbib}
\usepackage{amssymb}
\usepackage{float}

\usepackage{makeidx}
\usepackage{amsmath}
\usepackage{rotating}
\usepackage{longtable}
\usepackage{subfigure}
\usepackage{subfloat}
\usepackage{morefloats}
\usepackage{captcont}
\nocite{*}

\newcommand{\etal}{et~al.\/}

\newcommand{\kms}{\mbox{km\thinspace s$^{-1}$}}

\def\vlsr{v_{\rm LSR}}
\def\kms{\mathrm{km\,s^{-1}}}

\shortauthors{Werk \etal}
\shorttitle{Ionized Galactic Fountain}

\begin{document} 

\title{The Nature of Ionized Gas in the Milky Way Galactic Fountain}

\author{Jessica K.\ Werk\altaffilmark{1},
  K. H. R. Rubin \altaffilmark{2},
 H. V. Bish \altaffilmark{1},
J. X. Prochaska \altaffilmark{3,4},
Y. Zheng\altaffilmark{5, 10}, 
J. M. O'Meara\altaffilmark{6}, 
D. Lenz\altaffilmark{7, 8}, 
C. Hummels\altaffilmark{8}, 
A. J. Deason\altaffilmark{9}
}

\altaffiltext{1}{University of Washington, Seattle, WA $jwerk@uw.edu$}
\altaffiltext{2}{San Diego State University}
\altaffiltext{3}{UCO/Lick Observatory; University of California, Santa Cruz, CA }
\altaffiltext{4}{Kavli Institute for the Physics and Mathematics of the Universe (Kavli IPMU; WPI),
The University of Tokyo, Japan} 
\altaffiltext{5}{University of California, Berkeley, CA}
\altaffiltext{6}{W. M. Keck Observatory  65-1120 Mamalahoa Highway, Kamuela, HI, 96743}
\altaffiltext{7}{Jet Propulsion Laboratory, California Institute of Technology, Pasadena, California 91109, USA}
\altaffiltext{8}{California Institute of Technology, Pasadena, California, 91125 USA}
\altaffiltext{9}{Institute for Computational Cosmology, University of Durham, UK}
\altaffiltext{10}{Miller Institute for Basic Research in Science, UC Berkeley, CA}

\begin{abstract}

We address the spatial scale, ionization structure, mass and metal content of gas at the Milky Way disk-halo interface detected as absorption in the foreground of seven closely-spaced, high-latitude halo blue horizontal branch stars (BHBs) with heights $z = 3 - 14$~kpc.  We detect transitions that trace multiple ionization states (e.g. CaII, FeII, SiIV, CIV) with column densities that remain constant with height from the disk,  indicating that the gas most likely lies within $z < $ 3.4 kpc. The intermediate ionization state gas traced by CIV and SiIV is strongly correlated over the full range of transverse separations probed by our sightlines, indicating large, coherent structures greater than 1 kpc in size. The low ionization state material traced by CaII and FeII does not exhibit a correlation with either  N$_{\rm HI}$ or transverse separation, implying cloudlets  or clumpiness on scales less than 10 pc.  We find that the observed ratio log(N$_{\rm SiIV}$/ N$_{\rm CIV}$), with a median value of $-$0.69$\pm$0.04, is sensitive to the total carbon content of the ionized gas under the assumption of either photoionization or collisional ionization.  The only self-consistent solution for photoionized gas requires that Si be depleted onto dust by 0.35 dex relative to the solar Si/C ratio,  similar to the level of Si depletion in DLAs and in the Milky Way ISM. The allowed range of values for the areal mass infall rate of warm, ionized gas at the disk-halo interface is  0.0003 $<$ d$M_{\rm gas}$ / d$t$d$A$ [M$_{\odot}$ kpc$^{-2}$yr$^{-1}$] $<$ 0.006.  Our data support a physical scenario in which the Milky Way is fed by complex, multiphase processes at its disk-halo interface that involve kpc-scale ionized envelopes or streams containing pc-scale, cool clumps.

\end{abstract}


\section{Introduction}
\label{sec:intro}

The gaseous halo of the Milky Way supports a dynamic baryon cycle that provides fuel for future star formation and collects the by-products of stellar evolution \citep{putman12}.  This cycle is often envisioned as a Galactic fountain, in which hot, over-pressurized gas is ejected from the disk into the halo, and subsequently cools, recombines and rains back down onto the disk \citep{shapiro76, fraternali08}. However, the physical mechanisms that inject energy and material into the fountain and the detailed gas physics that follow are likely considerably more complex \citep[e.g][]{kim18}. Our ability to detect the gas flowing into and out of the Galaxy at small distances -- just as it emerges from the disk, or accretes to fuel the ISM -- presents a unique opportunity to constrain the physics that drive the Galactic fountain. 

Despite a rich observational history of absorption and emission studies, following the complete gas cycle of star-forming galaxies, from halo to disk, has remained a challenge. Since \cite{munch61} first observed absorption from CaII H $+$ K lines in the spectra of high-latitude early-type stars, it has been hypothesized that the Milky Way is enveloped in a diffuse, multiphase halo \citep{spitzer56}. More recently, absorption-line experiments using bright background sources, including halo stars and quasars, have been used to probe the diffuse circumgalactic gas surrounding our own Galaxy and others at projected distances $>$ 10 kpc \citep[e.g. ][]{spitzer65, bergeron86, savage94, lanzetta95, sembach03, fox04, lehner12, prochaska11}.  HI 21-cm emission maps of our own Galaxy's halo reveal giant arches, shells, and complexes of  neutral clouds moving with 25 $<  |$v$_{\rm LSR}| <  $~250 km/s \citep[e.g.][]{wakker97}. The Wisconsin H$\alpha$ Mapper (WHAM; Haffner et al. 1999\nocite{haffner99}) has found extended, diffuse Galactic  H$\alpha$ emission above the galactic plane, not associated with HII regions. Studies of the pulsar dispersion measure (DM) in the Galactic halo constrain the electron density of a warm-ionized medium (WIM) with a  characteristic height of 2 kpc above the disk midplane \citep{gaensler08}. The physical interpretation of these rich datasets is hampered by limited distance constraints, sparsely sampled background sightlines,  and degeneracies between local ``cloud" sizes and the unknown phase structure of the gas. 

Absorption-line studies of the Milky Way halo itself have used both distant QSOs and bright halo stars to constrain the nature and amount of diffuse, high velocity ($|v_{\rm LSR}|$ $>$ 90 km s$^{-1}$) material beyond the disk. The Galactic corona exhibits a range of ionization states traced by numerous ionized metal line transitions. From observations with FUSE, and later, COS, we see widespread, yet patchy, absorption from SiIV, CIV, and OVI that indicate T$\approx$10$^{5}$ K gas at scale heights of 2 $-$ 5 kpc \citep[e.g.][]{savage09}. Furthermore, ionized, diffuse, high-velocity gas is distributed over large spatial scales, with a covering fraction of $\sim$60\% \citep{savage03,sembach03}. Based on conservative ionization model assumptions, this gas provides enough potential material to sustain star-formation in the Milky Way for many Gyrs \citep{lehner11}.   

We present observations that directly constrain the kinematics, physical coherence scale, and mass and metal content of high latitude ionized gas at the Milky Way disk-halo interface. The disk-halo interface region of the Galaxy includes extraplanar, ionized material sometimes referred to as the WIM, thick disk, and corona, and roughly extends to $\pm$10 kpc above and below the disk. It additionally contains well-known structures mapped in HI 21-cm emission commonly referred to as intermediate and high velocity clouds (IVCs and HVCs).  Using a bracketing method that requires spectroscopy toward foreground and background halo stars with known distances, the heights to IVCs are constrained to lie within $\lesssim$3 kpc, and those of HVCs are mostly constrained to lie within heights $\lesssim$15 kpc \citep[e.g.][]{danly93, wakker96,wakker01, wakker07, thom08, wakker08}. 

In this study, we map the absorption signatures toward a sample of 7 halo blue horizontal branch stars (hereafter BHBs)  at different heights out to 14 kpc in the direction of the Northern Galactic Pole. To select our targets,  we used a catalog of $>$ 4500 BHBs with well-calibrated distances and velocities \citep{xue11}. The BHBs in our sample lie at high galactic latitudes, $b$ $>$ 60$^{\circ}$. At these latitudes, we minimize ambiguities in tracking inflowing and outflowing material via blueshifts and redshifts along the lines of sight. We also minimize the absorption due to the Milky Way ISM along each line of sight.   With a small sample of background probes at various scale heights, our goal is to place physical constraints on extraplanar ionized and neutral  material in the Milky Way. 

Throughout this manuscript we will report gas radial velocities, v$_{\rm rad}$,  along given lines of sight from our vantage point at the location of the sun in the Milky Way Galaxy. As we have no information about the transverse velocity of the gas perpendicular to the line of sight, it is important to distinguish these 1D velocities from true spatial velocities. Furthermore, we report all velocities in the frame of the local standard of rest (LSR), which accounts for the peculiar velocity of the sun with respect to the rotation of the galaxy.

\begin{deluxetable*}{cccccccccccc}
\tablewidth{0pc}
\tablecaption{BHB Star Properties\label{tab:bhbprops}}
\tabletypesize{\scriptsize}
\tablehead{\colhead{BHB Name, ID} & \colhead{RA} & \colhead{dec} & \colhead{$l$}  &\colhead{$b$} &\colhead{m$_{\rm FUV}$} &\colhead{m$_{\rm g}$} &\colhead{HRV} &\colhead{d} &
\colhead{x} & \colhead{y} & \colhead{z} \\
 & $(^\circ)$ &  $(^\circ)$ &  $(^\circ)$ & $(^\circ)$ &  & & (km s$^{-1}$) & (kpc) & (kpc) & (kpc) & (kpc)}
\startdata
J1527+4027 BHB1 &  231.9586 &   40.4580 &   65.6478 &   55.2262 & 16.30 & 13.78 & $-105.6$ $\pm$0.9 &   4.1 &   7.0 &  -2.1 &   3.4\\
J1534+5015 BHB2 &  233.5468 &   50.2657 &   80.9667 &   51.4485 & 16.95 & 14.20 & $-129.4$ $\pm$1.8 &   5.2 &   7.5 &  -3.2 &   4.1\\
J1344+1842 BHB3 &  206.0184 &   18.7164 &    0.8905 &   75.2642 & 17.27 & 14.46 & $-74.7$ $\pm$1.4 &   5.9 &   6.5 &   0.0 &   5.7\\
J1325+2232 BHB4 &  201.4767 &   22.5473 &    4.4384 &   80.9410 & 17.83 & 14.90 & $-100.3$ $\pm$1.8 &   7.4 &   6.8 &  -0.1 &   7.3\\
J1415+3716 BHB5 &  213.9881 &   37.2828 &   67.9148 &   69.5097 & 17.97 & 15.57 & $-134.0$ $\pm$1.2 &   9.9 &   6.7 &  -3.2 &   9.2\\
J1341+2823 BHB6 &  205.3327 &   28.3996 &   42.3913 &   78.8949 & 18.07 & 16.07 & $-124.8$ $\pm$2.2 &  11.3 &   6.4 &  -1.5 &  11.1\\
J1459+4128 BHB7 &  224.9749 &   41.4809 &   69.9352 &   60.0843 & 18.47 & 16.72 & $-183.1$ $\pm$4.4 &  15.3 &   5.4 &  -7.2 &  13.3\\
\enddata
\tablecomments{{}  Properties of target BHB stars based on SDSS DR8 SEGUE data (see Xue et al. 2011 for details); all coordinates are in epoch J2000. GALEX FUV magnitudes are determined by cross matching the closest point source detection to each BHB target.  } 
\end{deluxetable*}

In Section 2 we describe the data and observations used in this study. Section 3 presents our line-profile analysis of the absorption-line data.   In Section 4, we present our results based on the column density and velocity centroid measurements.  Section 5 examines how the ratio of SiIV/CIV constrains both photoionization and collisional ionization models of the high-latitude, intermediate-velocity gas we observe. In Section 6, we summarize our key results, and discuss the implied spatial scales of the observed clouds and/or interfaces in the context of physical models of the Milky Way Galactic Fountain, and finally discuss the implications of possibly dusty, ionized disk-halo interface material. 

\section{Data and Sample: Target Blue Horizontal Branch Stars}
We targeted BHBs in the halo of the Milky Way toward the North Galactic Pole extending 13 kpc above the disk into the halo. Our catalog of blue horizontal branch stars comes from Xue et al. (2011), who selected BHBs from SDSS DR8 stars with fiber spectra. Halo blue horizontal branch stars are good beacons for four reasons: (1) they are blue point sources and therefore have bright continua at UV and optical wavelengths; (2) their UV spectra are characterized by a fairly featureless bright continuum with broad Lyman series lines \citep{brown12} because they are quite metal poor; (3) they are approximate ``standard candles" since their distances can be estimated from multi-band photometry with an accuracy of ~10\% \citep{deason11};  and (4) many of them are moving with $|v_{\rm rad}|$ $>$ 100 km/s which allows us to cleanly separate any absorption features intrinsic to the star from the disk/halo interface gas (the BHB intrinsic stellar spectra do not exhibit any detectable SiIV or CIV, but do seem to show a modest amount of CI, SiII, FeII, and CaII; discussed in \S \ref{sec:wavecal}).

\begin{figure}[t!]
\begin{centering}
\hspace{-0.3in}
\includegraphics[width=1.2\linewidth]{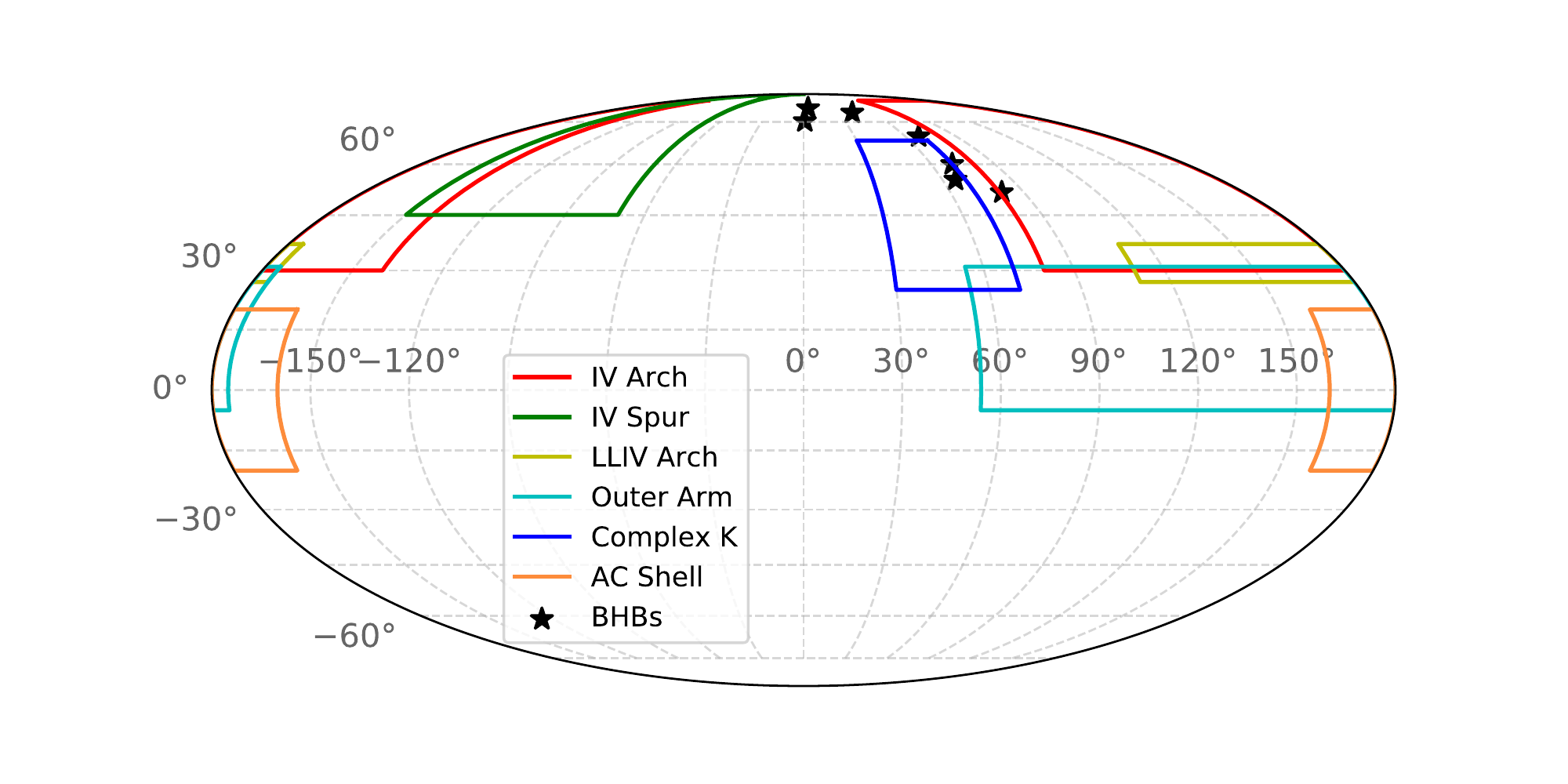}
\end{centering}
\caption{All-sky map in Galactic coordinates, with Galactic center at $l$$=$0$^\circ$, $b$$=$0$^\circ$, showing the positions of the 7 BHB sightlines as black stars. The colored shapes mark the approximate locations of well-studied intermediate-velocity (IV) complexes identified in HI 21-cm maps, with coordinates based on those given in \cite{ivcbook}. The positions of the BHBs skirt the very edge of the large, high-latitude structure known as the IV arch. For reference, Complex K has very patchy coverage in the area shown, velocities centered on v$_{\rm LSR}$ $\approx$ $-$80 km s$^{-1}$ which is higher than the absorption velocities we observe,  and much lower overall integrated HI column densities than the IV arch \citep{wakker01a}.}
\label{fig:map}
\end{figure}

The 7 BHBs in our COS sample have FUV magnitudes m$_{\rm FUV}$ $<$ 18.5, bright enough to be observed with COS in $<$ 8 orbits per star. All lie at a high Galactic latitude ($\ell$ $>$ 60$^{\circ}$), which in principle allows us to distinguish inflows from outflows above the Milky Way Disk. Table 1 presents the properties of our sample of 7 BHBs, with coordinates, distances, and stellar velocities. Reported BHB distances serve as an upper limit on the distance to any gas detected along the line of sight. Figure \ref{fig:map} shows our 7 sightlines on an all-sky map of the Milky Way, along with the approximate locations of well-studied intermediate-velocity extraplanar HI structures for reference. 

In Table 1, we additionally provide 3-dimensional Cartesian coordinates (given as xyz), commonly used in dynamical studies of the Milky Way halo.  This coordinate system is particularly useful for our purposes in examining the physical structure of gas in the frame of the Milky Way.  For reference, in this coordinate system, Galactic center is at the origin and the Sun lies at (x,y,z) = (8,0,0) kpc. The x-axis extends from Galactic center in the direction of the Sun, y is perpendicular to x in the plane of the disk, and z is perpendicular to the disk such that the system is right-handed, with positive z towards the North Galactic Pole. We will refer to z as height for the remainder of the manuscript.

\subsection{HST/COS Spectroscopy}
 
Far-UV spectra were obtained with the Cosmic Origins Spectrograph
\citep[COS;][]{froning09, green12} on the {\it Hubble Space Telescope} (Program ID:  14140). We observed our BHB targets with the COS G160M grating with a central wavelength of 1577 \AA~with exposure times adjusted to achieve a $S/N\approx 15$ per resolution element over $\lambda_{\rm obs} \approx 1385-1760$\AA. Table \ref{tab:bhbobs} contains a summary of the COS observations.

Our COS data probe absorption due to the following metal ions:  SiII ($\lambda$ 1526 \AA), FeII ($\lambda$ 1608 \AA), AlII ($\lambda$ 1670 \AA),  SiIV ($\lambda$ $\lambda$ 1393 \AA, 1402 \AA), and CIV ($\lambda$ $\lambda$1548 \AA, 1550 \AA). We refer to singly ionized species as low ions. These low ions have ionization potential energies ranging from $\approx$ 16 - 18 eV, and are most likely arising in cool gas photoionized by stellar radiation from the Milky Way, with some contribution from the extragalactic UV background (EUVB). We refer to SiIV and CIV as intermediate ions. The  ionization potential energies of SiIV and CIV are  of 45 eV and 64 eV, respectively. The intermediate ions may trace warm photoionized material, and/or may be produced in part by collisional ionization.

We combine the CALCOS-generated x1D files using v3.1.1 of the COADD\_X1D routine provided by the COS-GTO team \citep{danforth16}, which properly treats the error arrays of the input files using Poisson statistics. The code aligns the different exposures by determining a constant offset determined by cross-correlating strong ISM lines in a 10\AA~ wide region of the spectrum. With COS spectra, misalignments with magnitudes up to $\pm$30 km/s between spectral lines in the same target have been described, which leads to smearing and unphysical velocity offsets between different absorption lines. We discuss these wavelength calibration issues in detail in the next Section, \S~\ref{sec:wavecal}.

The COS line-spread function (LSF) is well described by
a Gaussian convolved with a power law that extends to many
tens of pixels beyond the line center \citep{coslsf}. 
These broad wings affect both the precision of our equivalent
width measurements and complicate assessments of line saturation.
We mediate these effects when we fit absorption lines
(described in Section 3) by convolving with the real LSF. Each
COS resolution element at R  $\sim$18,000 covers 16  km s$^{-1}$ and
is sampled by six raw pixels.  We performed our analysis on the data binned by three native spectral pixels to a dispersion of $\Delta\lambda$ $\approx$ 0.0367 \AA. The resulting science-grade spectra are characterized by  a FWHM $\approx$ 18 km/s.

\begin{deluxetable*}{llcccccc}
\tablewidth{0pc}
\tablecaption{Summary of Keck/HIRES and HST/COS G160M Observations\label{tab:bhbobs}}
\tabletypesize{\scriptsize}
\tablehead{\colhead{BHB Name  ID} & \colhead{HIRES: Date-Obs}  & \colhead{Exptime} &  \colhead{S/N}  & \colhead{decker, $\lambda$ Range}&\colhead{COS: Date-Obs} &\colhead{N$_{\rm orbit}$} &\colhead{S/N} \\
 &  & (s)  &   & ($\AA$) &  & &   }
\startdata
J1527+4027 BHB1 & 2015-May-14 &  400  & 48   & C1: 3824$-$6880 & 2015-Dec-29 &   1  &  14.1  \\
J1534+5015 BHB2 & 2015-Apr-12 & 420   &  31  &  C1: 3678$-$6576 & 2015-Dec-24 &   2 &  17.0 \\
J1344+1842 BHB3 & 2016-Mar-30 & 600   &  30 &   C5: 3830$-$8350& 2016-May-15 &   3 &  16.3 \\
J1325+2232 BHB4 & 2015-May-14 &  800  & 25   & C1: 3824$-$6880 & 2016-May-24 &   4 &  15.2  \\
J1415+3716 BHB5 & 2016-Mar-30 &  900  &  24  &   C5: 3830$-$8350 & 2016-Feb-21,27 & 6 & 16.0  \\
J1341+2823 BHB6 & 2015-May-14 &  1600  & 28    & C1: 3824$-$6880  & 2016-May-26, Jun-25 & 6  &15.7  \\
J1459+4128 BHB7 & 2017-Feb-20 & 1800    & 18   & C5: 3830$-$8350  & 2015-Dec-17,21&   8 & 15.3  \\
\enddata
\end{deluxetable*}

\subsection{Keck/HIRES Spectroscopy}
As a complement to our \emph{HST}/COS data, in 2015A we began obtaining Keck HIRES spectra covering CaII H$+$K ($\lambda\lambda$ 3934, 3969)  and NaI ($\lambda\lambda$ 5892, 5897) along the BHB lines of sight through several programs. The Keck HIRES data are fully described in \cite{bish19}, and are summarized here. Exposure times varied with the magnitude of the star, and are given in Table \ref{tab:bhbobs}. Generally, we aimed to achieve a S/N of $\gtrsim$20 over the range of 3900 - 5900 \AA. Our final wavelength ranges varied slightly with different cross-disperser and echelle angles for the BHBs on different nights, depending on the program for which it was observed.   We obtained HIRESr spectra with the C5 decker with the image rotator in vertical-angle mode, and the kv370 order blocking filter with 2 $\times$ 1 binning.  The C5 decker with a 1.148$\arcsec$ slit provides spectroscopic resolution of 1.3 km s$^{-1}$ pixel$^{-1}$ and R = 36,000, which corresponds to a velocity FWHM of 8.3 km s$^{-1}$. The C1 decker with a 0.861$\arcsec$ slit provides spectroscopic resolution of 1.0 km s$^{-1}$ pixel$^{-1}$ and R = 48,000, which corresponds to a velocity FWHM of 6.2 km s$^{-1}$.  Our final  velocity resolution of  5 - 10 km s$^{-1}$ allows us to distinguish material from the disk of the Milky Way and the star, and resolve different absorption components from each other. 

The ionization potential energies of CaII and NaI are 11.9 eV and 5.1 eV, respectively. Because they lie below 1 Ryd, these ions are likely associated with cool, predominantly neutral material.  However,  NaI is known to show up sporadically along Milky Way ISM lines of sight, and exhibit little correlation with gas phase metallicity or HI column density \citep{jenkins09}. Thus, the interpretation of the NaI line is complex, whereas CaII is a typical ion tracing neutral and low ionization-state gas. 

\subsection{Wavelength Calibration Errors in COS}
\label{sec:wavecal}

Here, we describe our analysis to ensure the COS wavelength solution of the co-added, aligned spectra produced by the Danforth routine \citep{danforth16} is accurate. Our strategy is to check our solution against CaII stellar absorption lines detected in Keck HIRES spectra and SDSS SEGUE stellar radial velocities of the same targets. We fit unsaturated, unblended stellar absorption lines of CaII in the Keck spectra and CI, SiII, and FeII in the COS spectra to check for potential systematic errors in the COS wavelength calibration that have been reported and discussed in detail by Wakker et al. (2015).  We fit stellar absorption profiles along with gas absorption profiles as described in \S \ref{sec:profs}. We compare our fitted stellar radial velocities from HIRES and COS spectra with the SDSS DR8 radial velocities for the BHBs in the v$_{\rm LSR}$ restframe.  
 
 Existing methods to test and correct for systematic errors in the COS wavelength calibration assume alignment in Milky Way interstellar absorption lines of various species, sometimes with a range of ionization potential energies \citep[e.g.][]{wakker15, danforth16}. The primary benefit of these methods is that they are broadly applicable to a variety of datasets, including QSO spectra. Instead, our method assumes stellar absorption lines of different ionic species observed with different instruments should give a consistent value for the stellar radial velocity of each BHB star.  Notably, our method will allow for velocity offsets between different ionic species due to foreground gas absorption at z$\approx$0  along each BHB line of sight. Thus, any differences between gas kinematics of different ionic species at the disk-halo interface of the Milky Way may correspond to physical velocity offsets between different gas phases.  

\begin{figure}[t!]
\begin{centering}
\hspace{-0.2in}
\includegraphics[width=1.1\linewidth]{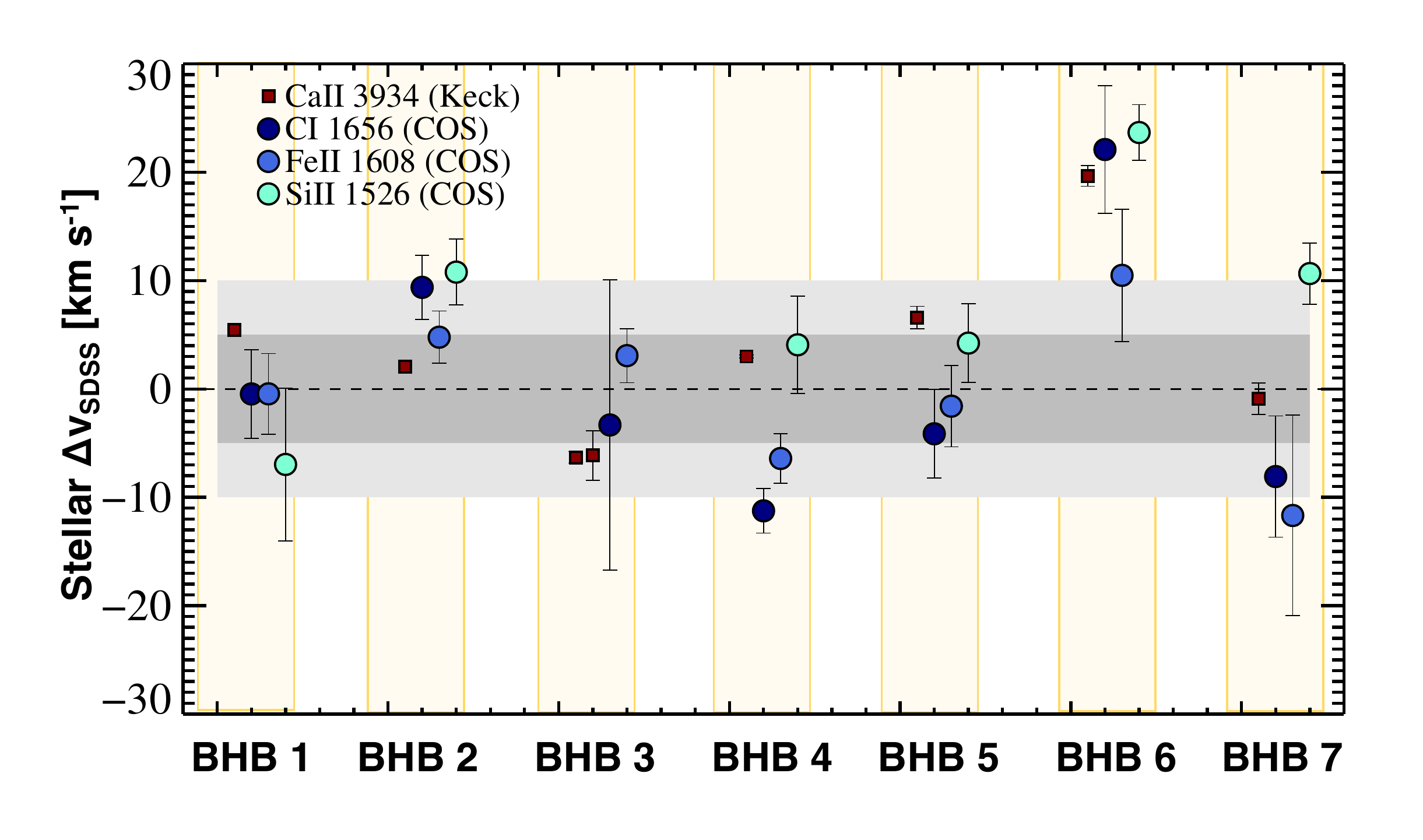}
\end{centering}
\caption{The difference between SEGUE BHB stellar radial velocities as measured in SDSS DR8 spectra and those we measure in our Keck/HIRES and {\emph{HST}}/COS spectra for different ionic transitions. The data are organized along the x-axis by BHB ID number, as given in Table 1, and by ionic transition as given in the key on the upper left of the plot. Stellar absorption is apparent in CaII $\lambda$3934 (Keck), CI $\lambda$1656, FeII $\lambda$ 1608, and SiII $\lambda$ 1526 (COS) for most of our lines of sight. The gray shaded regions indicate the approximate mean RMS of the dispersion solutions for our Keck HIRES (dark gray; $\sim$ 2 km s$^{-1}$) and COS G160M (light gray; $\sim$ 7.5 km s$^{-1}$) spectra.}
\label{fig:stellarvel}
\end{figure}

Figure \ref{fig:stellarvel} shows velocity offsets of stellar absorption lines between those measured in SDSS spectra \citep{xue11} and those measured in our Keck or COS spectra.  We show the offsets ($\Delta$v$_{\rm SDSS}$) for the different ionic species that consistently exhibit stellar absorption: CaII $\lambda$ 3934 (Keck), CI $\lambda$ 1656, FeII $\lambda$ 1608, and SiII $\lambda$ 1526 (COS).  We use a rest wavelength for CI $\lambda$ 1656 of 1656.9283, which is the wavelength of the $^{3}$P$_{0}$ ground state transition of CI, and do not include any contribution from weaker excited fine structure levels of the CI multiplet that span $\pm$335 km/s since they do not appear to be present in our data.   We note that AlII $\lambda$ 1670 often shows absorption from the BHB star, however we exclude it here because it is invariably saturated and often blended with the Milky Way interstellar absorption, which makes its fitted velocity centroid more uncertain. 

Figure \ref{fig:stellarvel} shows that for most BHBs, the stellar radial velocities measured in Keck and COS spectra are consistent with SDSS radial velocities within approximately $\pm$10 km s$^{-1}$. The velocity centroids of CaII 3934 as measured in the Keck data (small, red squares) are generally in better agreement with the SDSS radial velocities, with a median $|\Delta$v$_{\rm SDSS}|$ of 5.5 km s$^{-1}$. Next, we note that there are no apparent trends between $\Delta$v$_{\rm SDSS}$ and wavelength for the stellar absorption lines detected in the COS G160M spectra. The gray shaded regions indicate the approximate mean RMS of the dispersion solutions for our Keck HIRES (dark gray; $\sim$ 2 km s$^{-1}$) and COS G160M (light gray; $\sim$ 7.5 km s$^{-1}$) spectra. Reported errors in the SDSS heliocentric radial velocities for each BHB range between 1 and 3 km s$^{-1}$.

 In general, we find SDSS radial velocities are in agreement with those measured from the Keck and COS spectra. One exception is the stellar absorption we detect along the line of sight to BHB 6 (J1341+2823). The fits to the stellar velocities in both Keck HIRES and COS G160M spectra show $>$ 10 km s$^{-1}$ offset from the SDSS-reported heliocentric radial velocity (HRV). Despite this lack of agreement with SDSS, the HIRES and COS spectra give self-consistent results for BHB 6. The remainder of our analysis does not use SDSS spectra, so if there is an error in the SDSS spectra, it has no effect on our analysis. Overall, Figure \ref{fig:stellarvel} indicates that stellar radial velocities inferred from our HIRES and COS spectra are largely consistent within typical wavelength calibration errors for each instrument. Hidden systematic errors in the COS G160M wavelength solution that have been reported elsewhere (e.g Wakker et al. 2012) do not appear to significantly impact our BHB star COS spectra. 

\subsection{EBHIS HI 21cm Profiles}
HI 21-cm emission can provide constraints on the distribution of neutral hydrogen in the general vicinity of our BHB sightlines. However, direct comparisons of measured column densities in absorption and emission data are ill-advised for two reasons.  First, the spatial resolution of the HI data is 10.8', orders of magnitude larger than the much narrower $\sim$1$\arcsec$ HIRES slit or the 2.5$\arcsec$ circular COS aperture. This disparity implies that emission and absorption measurements do not trace material on the same spatial scales. Second, there are only radial velocity and no distance constraints for material detected in 21-cm emission. Nonetheless, these HI data are useful because of their all-sky coverage and their excellent constraints on the large-scale structure of the intermediate velocity material we wish to examine.

\begin{figure*}[t!]
\begin{centering}
\hspace{1.52in}
\includegraphics[width=0.65\linewidth]{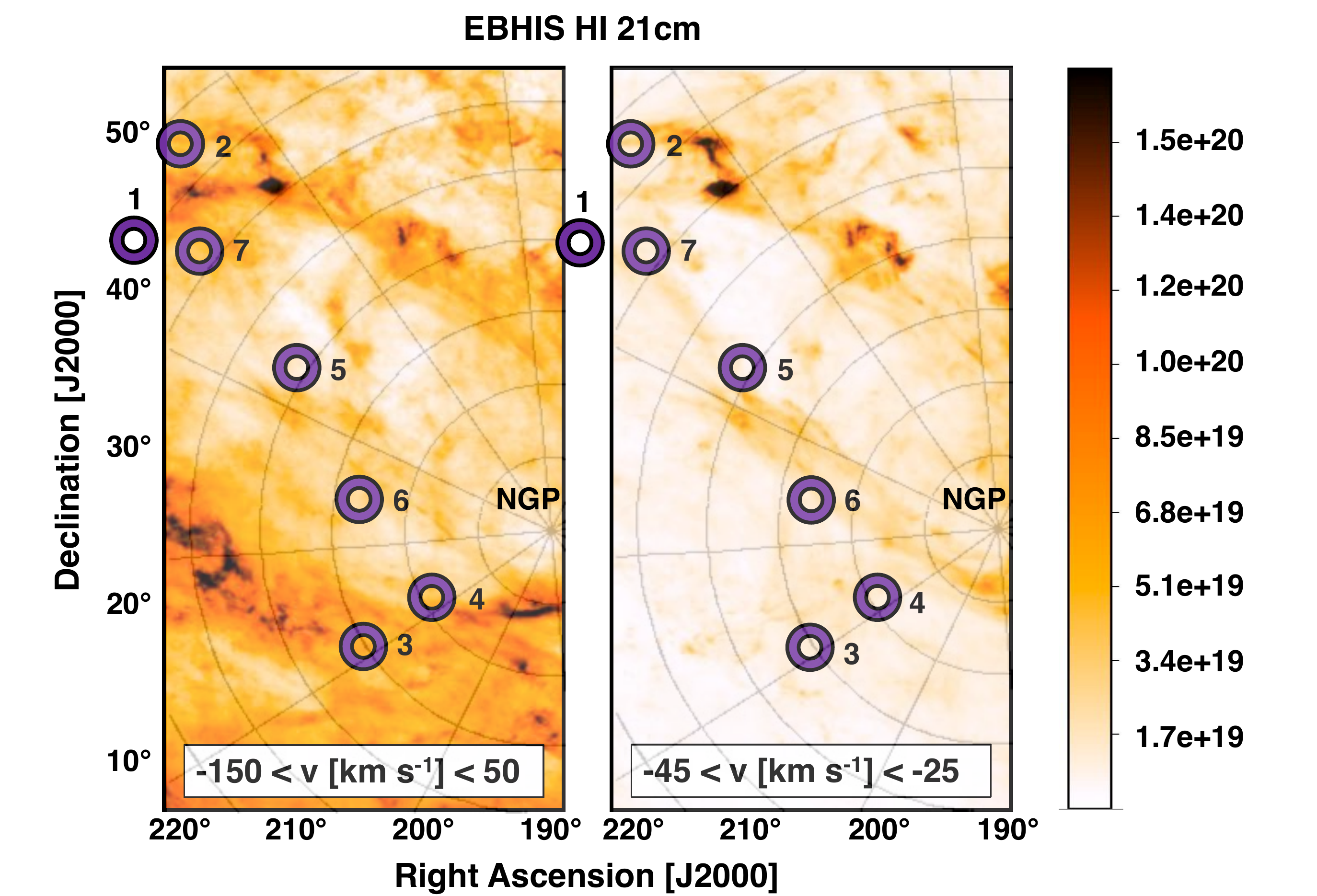}
\end{centering}
\caption{Integrated HI 21-cm column density maps from EBHIS (beam FWHM = 10.8$\arcmin$; Winkel et al. 2016) near the North Galactic Pole (NGP) in the vicinity of the 7 BHB targets for two velocity ranges: (left) a broad range, $-150~\mathrm{km~s^{-1}} < \vlsr < 50~\mathrm{km~s^{-1}}$  and (right) a narrower range, centered on the bulk of the intermediate velocity gas, $-45~\mathrm{km~s^{-1}}< \vlsr < -25~\mathrm{km~s^{-1}}$. The NGP is marked on each image, and the grid curves show rings of constant Galactic latitude ($b$) in 5$^{\circ}$ increments, while the diagonal lines show constant Galactic longitude ($l$) in 30$^{\circ}$ increments (BHBs 3 and 4 lie near the $l$ $\approx$ 0$^{\circ}$ line). Additionally, the approximate R.A. and Dec in degrees are given on the x and y axes, respectively. The location of each BHB is marked with a purple circle (unfortunately, BHB 1 lies just off the edge of the processed images). The high-latitude BHBs lie on the very low $l$ edge (0$^{\circ}$ $<$ $l$ $<$ 80$^{\circ}$) of the well-known IV-structure known as the ``IV-arch'' covering a large angular area on the sky, constrained along several lines of sight to have  0.4 kpc $<$ d $<$ 1.8 kpc (e.g. Wakker et al. 2001).}
\label{fig:HIImage}
\end{figure*}

We use data from the Effelsberg-Bonn HI Survey (EBHIS; Winkel et al. 2016\nocite{winkel16}), which images the full sky with a bandwidth of 100~MHz. The effective beam FWHM is 10.8$\arcmin$ and the effective velocity resolution is 1.44 km/s. We show in Figure \ref{fig:HIImage} integrated HI 21-cm column density maps from EBHIS over two velocity ranges (left: a broad velocity range $-150~\kms < \vlsr < 50~\kms$ ; right: the velocity range over which we detect the bulk of our IV absorption, $-45~\mathrm{km~s^{-1}}< \vlsr < -25~\mathrm{km~s^{-1}}$) in the approximate location of our sample of BHBs. Figure \ref{fig:HIImage} shows that none of our BHB lines of sight intersect the dense neutral clumps visible in this image, and instead seem to trace the wispy structures that extend across tens of degrees. 

At the position of each BHB, we download the HI spectra that are generated using a weighted interpolation with a Gaussian kernel from the HI Survey server of the Argelander-Institut f\"ur Astronomie\footnote{\url{https://www.astro.uni-bonn.de/hisurvey/AllSky$\_$profiles/}}. We show these EBHIS HI-21cm brightness temperature profiles as a function of $\vlsr$ on the right hand column of Figure 4. Intermediate velocity neutral gas with $\vlsr < -25~\kms$ is apparent along each of our seven lines of sight. In order to determine the HI column density in the IV gas along each line of sight, we perform simple,  dual Gaussian fits to these double-peaked, asymmetric profiles seen in Figure \ref{fig:exspeckeck}. In the optically thin limit, the HI column density is directly proportional to the integrated 21-cm emission line brightness (T$_{\rm b}$), which we calculate using the standard relation \citep{draine}:
\begin{equation}
    \frac{N_{\rm HI}}{\rm cm^{-2}} \approx 1.82\times10^{18} \int_{}^{} \left[\frac{T_b(v)}{\rm K}\right] d \left(\frac{v}{\rm km~s^{-1}}\right)
\end{equation}

We use the FWHM, velocity centroid, and the peak $T_b$ from the simple Gaussian fits to the IV components to determine their column densities within the 10.8$\arcmin$ beam. The relatively large beam-size gives the average N$_{\rm HI}$ on scales of 9.4 pc at a distance of z$=$3 kpc [or $\ell$ $\approx$ $(3.1$ pc)$\times$(z$/$1 kpc)]. Smaller-scale structures will be averaged out, including any unresolved structures probed by the absorption-line data along the BHB sightlines. Thus, values of N$_{\rm HI}$ determined from EBHIS data cannot be used as either upper or lower limits to the neutral gas column density associated with the absorption we see in Keck/HIRES or HST/COS spectra. However, the variation in N$_{\rm HI}$ over the area traced by our sightlines will prove useful for our analysis. 

\section{Line Profile Analysis}
\label{sec:profs}

\begin{figure*}[h!]
\begin{centering}
\includegraphics[width=0.99\linewidth]{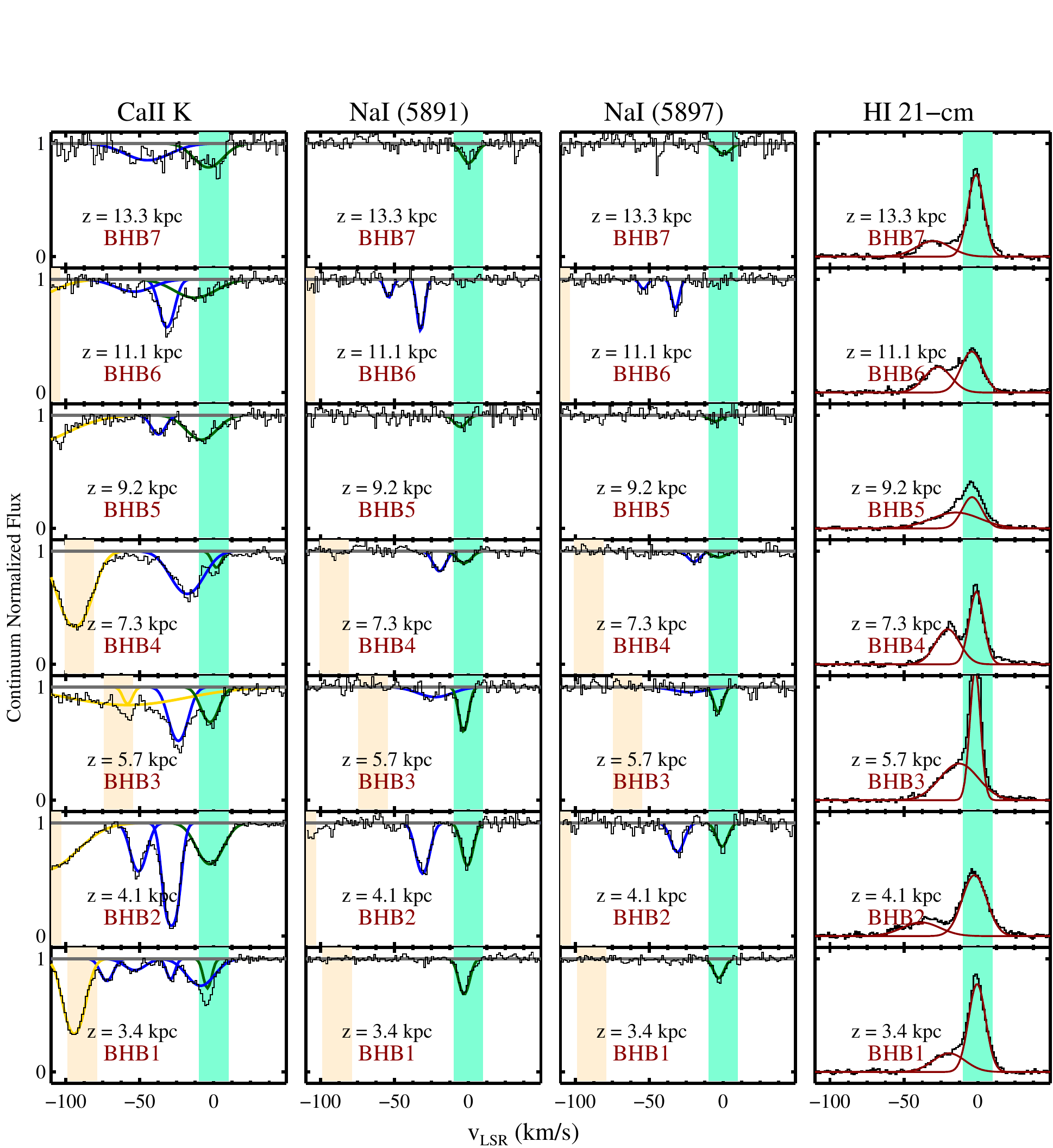}
\end{centering}
\caption{The first three columns show absorption line stacks of the NaI D doublet and CaII K drawn from the Keck/HIRES spectra. The velocities are relative to $\vlsr$, and the spectra have been continuum-normalized. The rightmost column shows the EBHIS HI 21-cm brightness temperature as a function of $\vlsr$. The peak brightness temperature, $T_{\rm b}$, on the y-axis of the rightmost column is $\approx$ 5.3 K. The 25 km/s-wide, green highlighted band centered on Galactic emission at $\vlsr=0~\kms$ matches the strongest HI feature in every case. In the Keck spectra, the beige highlighted band marks the radial velocity of the BHB as $\vlsr$ from SEGUE, for reference. We detect broad stellar absorption features in CaII, but not in NaI. The blueshifted features relative to the Milky Way ISM represent gas along the line of sight to the BHB, which has its height from the disk given on the lower left hand corner of each subplot. Our Voigt profile fits are shown in different colors on each plot. Features identified as IV absorption are shown in blue, while components we associate with the Milky Way ISM are shown in dark green. Two-component Gaussian fits to the HI brightness temperature plots are shown in deep red on the right-most panels. }
\label{fig:exspeckeck}
\end{figure*}

\begin{figure*}[t!]
\begin{centering}
\hspace{0.1in}
\includegraphics[width=0.95\linewidth]{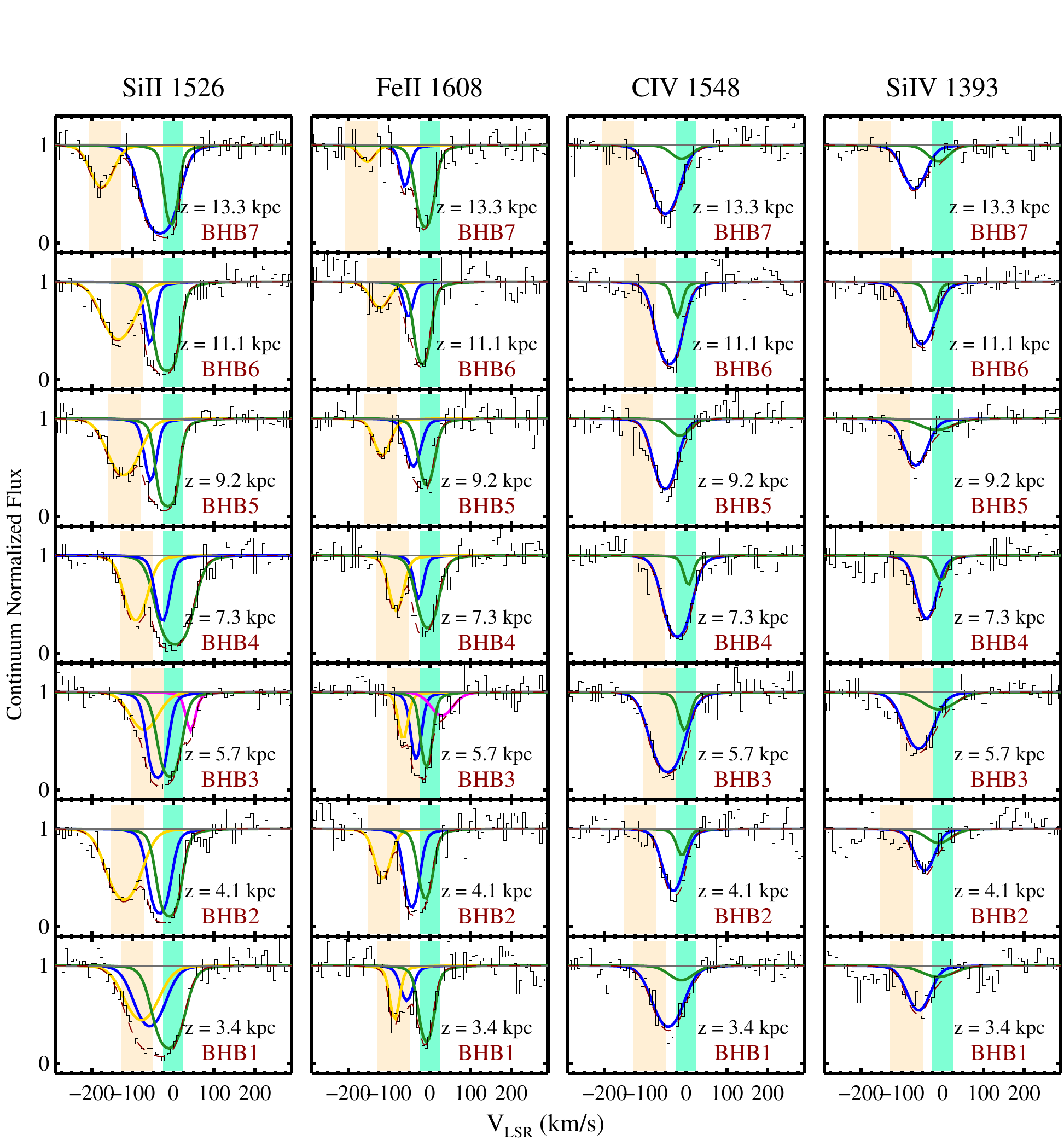}
\end{centering}
\caption{Ionic species drawn from the {\emph{HST}}/COS absorption-line spectra of the 7 target BHBs, centered on the transitions relevant to this study:  \ion{Si}{2}, \ion{Fe}{2}, \ion{C}{4}, and \ion{Si}{4}. The Voigt profile fits for individual components are shown in thick colored lines, while the total normalized line profile fit is shown in dashed red over the detected features. Blue lines highlight the IV absorption, gold lines highlight stellar absorption, and green lines show the Milky Way components. We show the SEGUE radial velocity of the BHB star in $\vlsr$, for reference, as a beige shaded area. Stellar absorption is present and fitted in the low ionization transitions of  \ion{Si}{2} and \ion{Fe}{2}, while it is absent from the intermediate ions. The central green shaded region is meant to guide the eye to $\vlsr=0~\kms$. The low ions in BHB3 distinctly show a weak redshifted component, highlighted in pink, although we do not discuss redshifted material in this work. }
\label{fig:exspec}
\end{figure*}

Lines of sight toward high Galactic latitude halo stars pass through Milky Way disk gas, extraplanar, ionized ``thick-disk" gas at the disk/halo interface, and the inner Milky Way CGM \citep{savage03, savage09, zheng18}. Figures \ref{fig:exspeckeck} and \ref{fig:exspec} show sections of the Keck/HIRES and HST/COS spectra that cover the ionized and neutral transitions we analyze in this work. These continuum-normalized spectra are given as a function of $\vlsr$, and the region we define as tracing the Milky Way ISM (henceforth ``MW") is highlighted in light green, $-20$ $<$ $\vlsr$ (km/s) $<$ 20 . The data themselves are shown as a black histogram, while the line-profile fits are shown as colored curves and are fully described later in this section.

We adopt terminology such that absorption features with LSR absolute velocities within $\sim 20 - 90~\kms$ of $\vlsr=0$ (km/s) are referred to as ``intermediate velocity" clouds (henceforth ``IV"). Features with absolute radial velocities $\vlsr >90~\kms$ are generally referred to as high velocity clouds (HVCs; Wakker et al. 2003\nocite{wakker03}), however we do not detect any HVCs in our dataset.  

Thus, our study focuses on extraplanar IV material, which must be understood within a rich context of previous work. Figure \ref{fig:map} shows our sightlines along with the locations of well-known IV structures identified via HI 21-cm maps. At a distance of $<$ 3.5 kpc \citep{ryans97}, from l $=$ 80$^\circ$ $-$ 220$^\circ$ and b $=$ 30$^\circ$ $-$ 80$^\circ$, lies the $\sim$10$^{5}$ M$_{\odot}$ IV arch, moving toward us at 40 - 90 $\kms$. This is the closest IV structure to our lines of sight, which lie just on its edge. Other less massive intermediate velocity structures include the IV Spur and the IV arch extension (LLIV), both of which exhibit scale heights $<$ 2 kpc \citep{wakker01a, wakker04}, and appear to be accreting at a mean velocity of $\approx -50~\kms$. For reference, lower and upper limits on the distances to most HVCs and IVCs typically place them in the inner (1 kpc $<$ d $<$ 10 kpc) Milky Way halo \citep{putman12}.

Along each halo star sightline, there is absorption from the Milky Way ISM immediately visible in low ionization state and neutral gas at approximately the local standard of rest velocity ($\vlsr = 0~\kms$) in its direction. Additionally, we often detect absorption lines of neutral and low-ion metals roughly coincident with the BHB star's radial velocity reported by SEGUE. The extent to which there are shifts between the centers of the beige bands and the line centroids of the stellar absorption in COS and Keck data highlights the uncertainties in the wavelength calibration that manifest as differences with the SEGUE stellar radial velocities in the LSR frame, and shown  in Figure 2. Together, the Milky Way ISM and stellar absorption do not account for the total observed absorption along each BHB line of sight. In the Keck/HIRES spectra, there are kinematically distinct absorption lines due to an additional intermediate velocity (IV) gas component(s). In the HST/COS spectra, the spectral resolution is too low to completely separate IV gas from Milky Way ISM, although we may assume both gaseous components are present as there is absorption over the full velocity range covered by the multiple components seen in the Keck/HIRES spectra. In this Section, we describe our approach to a Voigt profile fitting analysis with the goal of examining the extraplanar `IV' gas properties in both Keck/HIRES and HST/COS spectra. 

Before proceeding, we note two general caveats important for interpreting our results. First, in principal, distinguishing extraplanar and/or halo gas clouds from a coherent Milky Way ISM is accomplished by imposing a velocity cut in the LSR frame on absorption seen along the line-of-sight as described above. However, gas with velocities inconsistent with Milky Way differential rotation technically need not be spatially distinct from the MW ISM. Secondly, extraplanar material detected in absorption at coincident line-of-sight velocities may not itself be entirely co-spatial in a three-dimensional sense. This last caveat is one to which we will return when comparing the column densities of intermediate and low ionization transitions. 

Voigt profile fitting is an indispensable tool used to model and interpret absorption-line data, and is the best method for ``de-blending" stellar absorption, MW ISM, and IV gas. Generally, these profiles are generated using rules-based algorithms that make assumptions about the number of absorption components to fit, and sometimes demand correspondence between component structure in different ionic species. Each of these assumptions introduces a prior that can constrain parameter estimates in ways that are statistically valid, but do not capture the full physical uncertainty of the models. In particular, with COS, one has to worry about unresolved saturation effects, which are notorious for introducing specious column densities. Furthermore, higher spectral resolution tends to resolve more and more component structure, which is apparent  in visually comparing the higher-resolution HIRES with the lower-resolution COS absorption profiles. The spectral resolution and signal-to-noise of the data therefore set clear limits as to how many components one can fit and to what extent one can trust the derived fits. 

The standard approach to Voigt profile fitting is to fit as few components as required by the data for a reasonably good fit, with good being defined in a number of possible ways. In  \S \ref{secfit}, we detail all of the assumptions that go into our line-profile analysis. By adopting a consistent approach to Voigt profile fitting of absorption along each line of sight for every ion observed in our COS spectra we may understand the ionization structure of the material in the Milky Way's disk-halo interface. Specifically, we constrain the properties of the IV gas detected in the HST/COS spectra given a set of kinematic priors that come from higher-spectral-resolution Keck/HIRES spectra. 

\subsection{Qualitative Assessment and Overview} 

The FWHM of a single resolution element in our COS G160M spectra is more than two times broader than that of our HIRES spectra (18 km/s vs. 7 km/s). As a result, the Milky Way ISM is blended with absorption due to the IV extraplanar features we aim to examine in the COS data. The HIRES data allow a cleaner separation of Milky Way local ISM and additional absorption features, seen in Figure \ref{fig:exspeckeck}. The CaII K absorption at v$_{\rm LSR}$ $\approx$ 0 km/s lines up with the narrow, strong emission at v$_{\rm LSR}$ $\approx$ 0 km/s visible in the EBHIS HI-21cm data. Consistent with previous studies of  NaI that report its patchy detection in the Milky Way,  the v$_{\rm LSR}$ $\approx$ 0 km/s component of NaI absorption appears in only 4/7 cases (e.g. Meyer \& Lauroesch 1999\nocite{meyer99}). In addition, absorption consistent with the BHB stellar radial velocity is visible in CaII K, CI, SiII, AlII, and FeII (CaII H is catastrophically blended with broad stellar H$\epsilon$ absorption). We do not see stellar absorption in NaI, SiIV, and CIV in our HIRES or COS data for the targeted BHB stars.  

For all BHB stellar sightlines, a simple visual inspection of the line profiles for the ``intermediate ions," SiIV and CIV,  compared to those of the ``low ions", i.e. the neutral and singly ionized species, can instruct our profile fitting. In the COS spectra, the blended intermediate ions consistently exhibit broader line widths and more highly blueshifted velocity centroids ($\gtrsim$ 25 km s$^{-1}$) than the blended low ions. This apparent velocity offset most likely arises because the contribution of the predominantly neutral, higher-density, v$_{\rm LSR}$ $\approx$ 0 km/s, Milky Way ISM to the line profile is much greater in low ions than it is for intermediate ions. Adopting this assumption leads directly to the inference that extraplanar IV gas must exhibit a higher ionization parameter than the MW ISM.  The precise quantitative difference in ionization parameter (and density, by extension) between the MW and IV gas depends strongly on the total HI column of each component, the gas-phase metallicity of each component, and the background ionizing spectrum affecting each component. We note that with our current dataset, we cannot independently constrain any of these quantities. 

\subsection{A Fitting Procedure to Isolate Intermediate-Velocity Gas from Milky Way Absorption}
\label{secfit}

Our iterative fitting program for absorption lines detected with both Keck/HIRES and COS uses the \verb1MPFIT1 software\footnote{http://cow.physics.wisc.edu/$\sim$craigm/idl/fitting.html}\citep{mpfit}. The line profiles we derive from Voigt profile fitting of the COS data are convolved with the COS LSF at LP3 as given at the nearest observed wavelength grid point in the compilation by \cite{coslsf}. Different transitions of the same ionic species (e.g SiIV $\lambda$1393 and SiIV $\lambda$ 1402) are required to have the same component structure, and are therefore fit simultaneously to give a single solution. Each fit gives a value of the component column density $N$, Doppler $b$, and velocity centroid, $v$, in the LSR frame. The best fits emerge after a $\chi^{2}$ minimization process implemented in our code and depend strongly on the number of components assumed to be present. Reported errors in fitted parameters and overall goodness of fit via reduced $\chi^{2}$ are not independent of the assumptions in the analysis nor the limitations of the data (e.g. due to saturation).  

For the Keck/HIRES line profile analysis, we adopt a minimum number of components that results in profile fits to the data such that the reduced $\chi$$_{\nu}$$^{2}$ $<$ 1.1. Not including stellar absorption, we fit an average of 2$-$3 CaII components per sightline in the Keck data. Stellar absorption is seen in CaII (yellow fits shown in Figure \ref{fig:exspeckeck}), and is generally well-separated from the MW and IV gas. Each sightline has a distinctly separable MW component visible in CaII, sometimes detected in NaI. When NaI is detected in either IV or MW gas, its components' velocity alignment with the CaII components is excellent. 

In our analysis of the COS spectra, we proceed under the assumption that {\emph{both}} MW ISM and IV components are present with varying contributions in the low and intermediate ions. The main reason for ``forcing" a two-component fit is to be able to compare low and intermediate ionic column densities directly without worrying about different levels of contamination from a MW component. This assumption is physically motivated, and supported by the Keck/HIRES spectra which show a distinct MW component in CaII and HI along all sightlines. To ensure that two components are present in the COS data, our fitting procedure imposes lower limits on the column density of the MW components (details below). If these limits are not imposed, the iterative Voigt profile fits sometimes converge on a solution such that a single component dominates the result (i.e. the minimized second component contributes a column $<$ 10$^{9}$ cm$^{-2}$).  In general, the precise values of our imposed lower limits, as long as they are not unreasonably large so as to be completely inconsistent with the data, have very little impact on the final fit parameters. Finally, we note that our results, which focus primarily on IV SiIV and CIV, depend very little on the fitting assumptions described in this Section. 

 We present the component fits for each line of sight for select transitions in both Keck and COS spectra in Figures \ref{fig:exspeckeck} and \ref{fig:exspec}. We tabulate our component fitting results for the column density $N$, Doppler parameter $b$, velocity offset $v$ from $\vlsr$, along with associated errors in Tables \ref{tab:bhbmwfits} and \ref{tab:bhbivfits}, for the MW and IV components, respectively. We also show the results from the dual Gaussian fits to the HI 21-cm $T_{\rm b}$ profiles; in this case the $b$ column is in fact a Gaussian $\sigma$ and not a Doppler $b$ parameter. The Voigt profile fits give typical values of $\chi^2$ ranging from 200 $-$ 400, corresponding to values of reduced $\chi_{\nu}^2 < 1.4$. Our best constraints on the MW low-ion column densities in the COS data are from FeII, which is the least saturated low ion. Fits for MW CIV and SiIV are poorly constrained by the data, and thus, their best-fit parameters are often given as limits. Below, we describe our COS absorption-line fitting procedure and all associated assumptions in detail.

\begin{deluxetable}{ccccc}
\tablewidth{0pc}
\tablecaption{BHB Milky Way ISM Component Fits\label{tab:bhbmwfits}}
\tabletypesize{\scriptsize}
\tablehead{\colhead{BHB ID} & \colhead{Ion} & \colhead{log N$_{\rm ion}$} & \colhead{v$_{\rm LSR}$} & \colhead{$b$}}\\
\startdata
 & & & & \\
BHB1 & CaII& 11.48$\pm$0.06 &  -4.24$\pm$0.20 &   3.84$\pm$0.36 \\
&FeII& 14.57$\pm$0.08 & -10.30$\pm$2.27 &  16.36$\pm$3.50 \\
&SiII& 14.50$\pm$0.16 & $<$-14.24 &  35.43$\pm$3.78 \\
&AlII& 13.33$\pm$0.11 & -12.41$\pm$3.24 &  22.30$\pm$4.72 \\
&SiIV&$<$ 12.80 & $<$-14.24 & $>$ 50.00 \\
&CIV&$<$ 13.20 & $<$-14.24 &  37.06$\pm$21.44 \\
&HI& 19.97$\pm$0.01& -0.49$\pm$0.10&  8.16$\pm$0.14\\
 & & & & \\
\hline
 & & & & \\
BHB2 & CaII& 12.14$\pm$0.01 &  -2.65$\pm$0.19 &  11.64$\pm$0.28 \\
&FeII& 14.44$\pm$0.07 & $<$-12.65 &  16.48$\pm$2.73 \\
&SiII& 14.65$\pm$0.30 & $<$-12.65 &  22.33$\pm$3.86 \\
&AlII& 13.81$\pm$1.75 & $<$-12.65 &  12.83$\pm$11.73 \\
&SiIV&$<$ 12.80 & $<$-12.65 &  37.25$\pm$20.33 \\
&CIV&$<$ 13.20 & $<$-12.65 &  13.05$\pm$6.84 \\
&HI& 19.97$\pm$0.01& -2.30$\pm$0.14&  11.72$\pm$0.20\\
 & & & & \\
\hline
 & & & & \\
BHB3 & CaII& 11.89$\pm$0.00 &  -2.60$\pm$0.37 &   8.00$\pm$0.42 \\
&FeII& 14.64$\pm$0.38 & -10.39$\pm$3.57 & $<$ 10.00 \\
&SiII&$<$ 14.50 & $<$-12.60 &  23.72$\pm$5.05 \\
&AlII& 13.27$\pm$0.37 & $<$-12.60 &  15.19$\pm$8.21 \\
&SiIV&$<$ 12.80 & $<$-12.60 & $>$ 50.00 \\
&CIV& 13.42$\pm$0.34 &  -6.80$\pm$4.39 &  13.67$\pm$9.68 \\
&HI& 19.94$\pm$0.01& -1.78$\pm$0.04&  5.09$\pm$0.08\\
 & & & & \\
\hline
 & & & & \\
BHB4 & CaII& 11.25$\pm$0.08 &   1.92$\pm$0.43 &   4.38$\pm$0.67 \\
&FeII& 14.57$\pm$0.06 & $<$ -8.08 &  26.22$\pm$3.75 \\
&SiII& 14.68$\pm$0.22 &   0.85$\pm$13.57 &  38.13$\pm$7.83 \\
&AlII&$<$ 13.50 & $<$ -8.08 &  28.61$\pm$2.47 \\
&SiIV&$<$ 12.80 &  -7.06$\pm$3.81 &  12.88$\pm$7.16 \\
&CIV&$<$ 13.20 &   2.91$\pm$5.31 & $<$ 10.00 \\
&HI& 19.82$\pm$0.01& -0.90$\pm$0.11&  7.01$\pm$0.14\\
 & & & & \\
\hline
 & & & & \\
BHB5 & CaII& 11.94$\pm$0.03 &  -8.31$\pm$0.81 &  13.77$\pm$1.19 \\
&FeII&$<$ 14.80 &  -9.96$\pm$2.22 &  18.52$\pm$4.09 \\
&SiII& 15.17$\pm$0.83 & -16.57$\pm$3.84 &  15.67$\pm$6.07 \\
&AlII& 14.46$\pm$4.24 & $<$-18.31 &  13.76$\pm$18.28 \\
&SiIV&$<$ 12.80 & $<$-18.31 & $>$ 50.00 \\
&CIV&$<$ 13.20 & $<$-18.31 &  29.05$\pm$17.77 \\
&HI& 19.59$\pm$0.02& -3.91$\pm$0.19&  9.55$\pm$0.35\\
 & & & & \\
\hline
 & & & & \\
BHB6 & CaII& 11.95$\pm$0.08 & -12.74$\pm$2.89 &  19.12$\pm$3.41 \\
&FeII&$<$ 14.80 & -20.31$\pm$2.35 &  15.40$\pm$2.36 \\
&SiII& 14.85$\pm$0.24 & -18.64$\pm$3.14 &  21.21$\pm$4.44 \\
&AlII& 13.42$\pm$0.14 & -22.30$\pm$7.58 &  25.99$\pm$6.21 \\
&SiIV& 12.87$\pm$0.19 & $<$-22.74 & $<$ 10.00 \\
&CIV& 13.27$\pm$0.37 & -22.68$\pm$7.35 & $<$ 10.00 \\
&HI& 19.74$\pm$0.01& -4.15$\pm$0.28&  10.35$\pm$0.34\\
 & & & & \\
\hline
 & & & & \\
BHB7 & CaII& 11.92$\pm$0.04 &  -3.63$\pm$1.09 &  13.33$\pm$1.54 \\
&FeII& 14.73$\pm$0.12 & $<$-13.63 &  18.58$\pm$3.22 \\
&SiII& 14.85$\pm$1.16 &  -7.46$\pm$6.21 & $<$ 10.00 \\
&AlII& 15.50$\pm$0.46 & $<$-13.63 &  10.44$\pm$1.86 \\
&SiIV&$<$ 12.80 & $<$-13.63 & $>$ 50.00 \\
&CIV&$<$ 13.20 & $<$-13.63 & $>$ 50.00 \\
&HI& 19.92$\pm$0.01& -1.26$\pm$0.09&  7.78$\pm$0.13\\
\enddata
\tablecomments{{}  Voigt profile fit parameters for the Milky Way ISM
  components for a variety of ionic species observed along each BHB
  sightline.  The parameters are:  column density as log N
  (cm$^{-2}$), velocity centroid, v$_{\rm LSR}$ (km s$^{-1}$), and
  Doppler parameter, $b$, which quantifies the line width in km
  s$^{-1}$. We also tabulate the results from the Gaussian fits to the HI 21-cm T$_{\rm b}$ profiles; in this case the $b$ column is given as the Gaussian $\sigma$ $\times$ $\sqrt{2}$.  } 
\vspace{0.3in}
\end{deluxetable}

\begin{deluxetable}{ccccc}
\tablewidth{0pc}
\tablecaption{BHB Intermediate Velocity Component Fits\label{tab:bhbivfits}}
\tabletypesize{\scriptsize}
\tablehead{\colhead{BHB ID} & \colhead{Ion} & \colhead{log N$_{\rm ion}$} & \colhead{v$_{\rm LSR}$} & \colhead{$b$}}\\
\startdata
 & & & & \\
BHB1 & CaII& 11.95$\pm$0.02 &  -8.85$\pm$0.55 &  12.58$\pm$0.66 \\
&CaII& 11.30$\pm$0.04 & -29.11$\pm$0.24 &   3.87$\pm$0.36 \\
&CaII& 11.43$\pm$0.04 & -53.27$\pm$0.76 &   9.84$\pm$1.19 \\
&CaII& 11.52$\pm$0.03 & -72.30$\pm$0.30 &   5.90$\pm$0.44 \\
&FeII& 13.89$\pm$0.27 & -58.63$\pm$7.61 &  $16.20\substack{+17.42 \\ -15.20}$\\
&SiII&$<$ 14.22 & $>$-60.59 & $>$ 50.00 \\
&AlII& 12.98$\pm$0.33 & -59.41$\pm$2.89 &  10.05$\pm$7.20 \\
&SiIV& 13.35$\pm$0.04 & -61.80$\pm$2.75 &  31.39$\pm$4.18 \\
&CIV& 14.04$\pm$0.03 & -46.75$\pm$2.71 &  41.39$\pm$3.03 \\
&HI& 19.57$\pm$0.04&-19.45$\pm$1.11&  15.34$\pm$1.30\\
 & & & & \\
\hline
 & & & & \\
BHB2 & CaII& 12.56$\pm$0.01 & -28.60$\pm$0.05 &   5.86$\pm$0.06 \\
&CaII& 12.02$\pm$0.01 & -50.83$\pm$0.13 &   7.14$\pm$0.17 \\
&FeII& 14.62$\pm$0.09 & -46.07$\pm$2.42 &  15.49$\pm$3.34 \\
&SiII&$<$ 14.50 & $>$-36.40 &  22.30$\pm$21.77 \\
&AlII&$<$ 13.34 & $>$-38.84 &  25.39$\pm$24.84 \\
&SiIV& 13.26$\pm$0.04 & -46.91$\pm$3.67 &  26.27$\pm$4.37 \\
&CIV& 13.92$\pm$0.03 & -34.19$\pm$1.60 &  27.07$\pm$2.21 \\
&HI& 19.51$\pm$0.03&-38.40$\pm$0.75& 18.12$\pm$1.16\\
 & & & & \\
\hline
 & & & & \\
BHB3 & CaII& 12.14$\pm$0.02 & -23.96$\pm$0.23 &   8.20$\pm$0.37 \\
&FeII& 14.41$\pm$0.26 & -35.92$\pm$3.51 & $<$ 10.00 \\
&SiII& 14.65$\pm$0.53 & -41.45$\pm$10.73 &  17.91$\pm$14.57 \\
&AlII&$<$ 14.06 & $>$-42.09 & $<$ 10.00 \\
&SiIV& 13.58$\pm$0.02 & -61.84$\pm$2.09 &  39.11$\pm$3.07 \\
&CIV& 14.30$\pm$0.05 & -48.63$\pm$4.44 &  40.57$\pm$3.82 \\
&HI& 19.94$\pm$0.01&-12.62$\pm$0.47& 18.34$\pm$0.42\\
 & & & & \\
\hline
 & & & & \\
BHB4 & CaII& 12.25$\pm$0.01 & -17.97$\pm$0.33 &  14.36$\pm$0.49 \\
&FeII& 13.95$\pm$0.32 & -29.32$\pm$6.49 & $<$ 10.00 \\
&SiII& 14.05$\pm$0.64 & $>$-28.92 &  15.98$\pm$13.65 \\
&AlII&$<$ 13.41 & $>$-22.05 &  16.09$\pm$24.48 \\
&SiIV& 13.54$\pm$0.03 & -42.39$\pm$1.77 &  23.98$\pm$2.90 \\
&CIV& 14.27$\pm$0.03 & -24.32$\pm$1.80 &  32.37$\pm$2.05 \\
&HI& 19.73$\pm$0.02&-20.40$\pm$0.31&  11.71$\pm$0.47\\
 & & & & \\
\hline
 & & & & \\
BHB5 & CaII& 11.51$\pm$0.06 & -37.40$\pm$0.73 &   6.57$\pm$1.04 \\
&FeII& 14.12$\pm$0.12 & -42.46$\pm$4.24 &  18.75$\pm$6.56 \\
&SiII& 13.97$\pm$0.28 & -58.31$\pm$7.22 &  12.88$\pm$8.44 \\
&AlII& 13.46$\pm$0.41 & -35.50$\pm$13.82 &  28.80$\pm$6.80 \\
&SiIV& 13.38$\pm$0.03 & -68.77$\pm$2.10 &  31.21$\pm$3.18 \\
&CIV& 14.07$\pm$0.02 & -53.98$\pm$1.91 &  29.28$\pm$2.40 \\
&HI& 19.69$\pm$0.01&-14.83$\pm$1.02& 23.52$\pm$0.76\\
 & & & & \\
\hline
 & & & & \\
BHB6 & CaII& 11.98$\pm$0.06 & -31.69$\pm$0.32 &   6.59$\pm$0.55 \\
&CaII& 11.72$\pm$0.09 & -53.88$\pm$2.75 &  17.50$\pm$3.78 \\
&FeII& 13.81$\pm$0.19 & -54.75$\pm$5.47 & $<$ 10.00 \\
&SiII& 13.98$\pm$0.22 & -59.95$\pm$3.23 & $<$ 10.00 \\
&AlII& 12.97$\pm$0.69 & -61.79$\pm$5.87 & $<$ 10.00 \\
&SiIV& 13.61$\pm$0.04 & -54.30$\pm$2.46 &  34.11$\pm$2.08 \\
&CIV& 14.27$\pm$0.04 & -44.00$\pm$1.92 &  29.75$\pm$1.85 \\
&HI& 19.63$\pm$0.02&-26.95$\pm$0.52&  12.73$\pm$0.68\\
 & & & & \\
\hline
 & & & & \\
BHB7 & CaII& 11.94$\pm$0.04 & -44.74$\pm$2.12 &  21.32$\pm$0.97 \\
&FeII& 13.93$\pm$0.11 & -63.15$\pm$3.01 &  10.27$\pm$5.44 \\
&SiII& 14.66$\pm$0.06 & -35.81$\pm$2.57 &  39.64$\pm$2.23 \\
&AlII& 12.99$\pm$0.07 & -64.94$\pm$3.74 &  19.82$\pm$5.69 \\
&SiIV& 13.34$\pm$0.03 & -71.76$\pm$2.08 &  31.10$\pm$3.21 \\
&CIV& 14.10$\pm$0.02 & -53.76$\pm$1.56 &  36.50$\pm$2.26 \\
&HI& 19.52$\pm$0.03&-30.00$\pm$0.67& 16.38$\pm$1.06\\
\enddata
\tablecomments{{}  Voigt profile fit parameters for the intermediate velocity components for a variety of ionic species observed along each BHB sightline.  The parameters are: column density as log N (cm$^{-2}$), velocity centroid, v$_{\rm LSR}$ (km s$^{-1}$), and Doppler parameter, $b$, which quantifies the line width in km s$^{-1}$. We also tabulate the results from the Gaussian fits to the IV component of the HI 21-cm T$_{\rm b}$ profiles; in this case the $b$ column is given as the Gaussian $\sigma$ $\times$ $\sqrt{2}$. } 
\end{deluxetable}

\begin{itemize}
\item{{\bf{Identifying the Stellar Absorption by SEGUE Radial Velocity:}} As described in \S \ref{sec:wavecal}, the low-ion transitions in COS show absorption roughly consistent with SEGUE-reported stellar radial velocities. When present, we fit this stellar component allowing N, $b$, and v$_{\rm cen}$ to vary freely, and the results are consistent with expectations from HIRES spectra in which stellar absorption in CaII is cleanly separated from the MW and IV components. In Figures \ref{fig:exspeckeck} and \ref{fig:exspec}, the stellar component is colored in gold. In one case, BHB3, absorption centered on the expected stellar velocity in CaII demands two components, one very broad, and one weak, narrow component. Physically, both may not be associated with the star, but we proceed by simply noting that that the IV column density for CaII along this sightline may be an underestimate. }
\item{{\bf{Defining the MW Component in Velocity Space: }} For every BHB line of sight observed with COS, we fit one Milky Way ISM component within $\pm$10 km s$^{-1}$ of the LSR velocity for the CaII K MW line fit as defined by the HIRES spectra (v$_{\rm cen, CaII}$). We force this MW component to have a minimum column density as defined below.  In this analysis, we do not use the NaI line directly, since a MW component is not consistently detected in NaI due to its rather dramatic patchiness in the ISM (e.g. M{\"{u}}nch \& Zirin, 1961). However, we note that when NaI is detected in the HIRES spectra at v$_{\rm LSR}$ $\approx$ 0 km s$^{-1}$,  its velocity centroid is consistent with CaII MW absorption within a few km s$^{-1}$. In Figures \ref{fig:exspeckeck} and \ref{fig:exspec}, the MW component is colored in green. Table \ref{tab:bhbmwfits} shows that in approximately 50\% of the cases, our MW Voigt profile fits do not fully converge because they they are constrained not to venture outside of the $\pm$10 km s$^{-1}$ band.  In these cases, the values for the velocity centroid of the MW component are reported as limits: e.g. v$_{\rm cen}$ $<$ v$_{\rm cen, CaII}$ - 10 km s$^{-1}$. }
\item{{\bf{Lower Limit on the Low-Ion MW Component Column Densities:}}  The combined IV and MW components of the strong low-ion transitions detected in the COS spectra (SiII, AlII, FeII) show signs of saturation along every line of sight. Due to its non-Gaussian LSF, COS is known to saturate at normalized fluxes of $\sim$0.2. In our fitting procedure, we assume that the MW ISM dominates the column density contribution to the observed {\emph{saturated}} absorption features near v$_{\rm LSR}$ $\approx$ 0 km s$^{-1}$. This assumption corresponds to a minimum column density contribution from each transition which we can compute using its oscillator strength and rest wavelength, and assuming the lines are unresolved, with b$\approx$ 20 km s$^{-1}$. To compute this lower limit on the MW column density, we use the XIDL routine ``ism\_pltlin.pro."\footnote{https://github.com/profxj/xidl/blob/master/ISM/ism\_pltlin.pro} Lower-limit Log N [cm$^{-2}$] values for SiII, FeII, and AlII, respectively, are determined to be 14.5, 14.8, and 13.3. In practice, this limit serves to demand a dominant MW component for the neutral, low ionization state ISM.  Our minimization procedure converges to this MW column density lower limit for the low ions in only three cases seen in Table \ref{tab:bhbmwfits}.  Because the Voigt profile fits would prefer a lower column density than we allow, we report log N$_{\rm ion}$ in these three cases as conservative upper limits.  }
\item{{\bf{Setting Limits on the Intermediate-Ion MW Component Column Densities:}} Our primary assumption is that there are two {\emph{detected}} components along the BHB lines of sight corresponding to MW ISM and IV gas in the halo.  For the intermediate ions which do not have a dominant MW component visible in the data, we use the XIDL routine ``ism\_pltlin.pro" to determine the column density floors for SiIV and CIV. The median equivalent width 2$\sigma$ detection limits computed on each COS spectrum are 36 and 51 m$\AA$ for SiIV, and CIV, respectively.  Corresponding log N [cm$^{-2}$] values for SiIV and CIV are determined to be 12.8, and 13.2.  We artificially impose the requirement that the column densities not go below these upper limits when  we fit for two velocity components. In practice, this limit serves to demand a detected but weak MW component for the ISM in intermediate ionization states. The fits for the majority of the sightlines hit this lower column density limit in the intermediate ionization species 
because there are simply not easily separable MW components visible in the data. Indeed, most of the SiIV and CIV prefers a single component fit that is very similar to the reported IV fits in Table \ref{tab:bhbivfits}, with a slightly higher column density. Nonetheless, to present a consistent analysis, we demand a MW component in SiIV and CIV with the above minimum columns. Thus, in the tables, the reported fits are given as {\emph{upper}} limits, since the Voigt profiles would prefer to converge on an arbitrarily low column density for the MW intermediate ions.   }
\end{itemize}

The above steps define the primary assumptions that set our best two-component IV $+$ MW  fits in the COS spectra. In addition, we assume that each component in the COS spectra cannot have a Doppler $b$ parameter $<$ 10 km s$^{-1}$ or $>$ 50 km s$^{-1}$. These limits ensure that the iterative fitting procedure does not simply minimize a second component by giving it a very small or large line width. We define a lower limit of 10 km s$^{-1}$ to be the narrowest resolvable line given the COS line FWHM of 18 km s$^{-1}$.  For reference, the relation between the two quantities is given by: FWHM $\approx$ 1.665$b$. The precise values of the imposed Doppler $b$ parameter limits make very little difference to the IV and MW column density values, and the cases in which our profile fits hit these limits are mostly for the poorly-constrained MW components.  In the following Sections, we primarily focus on the properties of the IV component. In general this procedure produces well-constrained column densities and velocities for the IV components tabulated in Table \ref{tab:bhbivfits}, which are the focus of this study. 


\section{Results}

\subsection{Lack of Trends with Height from the Disk}
\label{sec:scaleheight}

\begin{figure*}[t!]
\begin{centering}
\hspace{0.40in}
\includegraphics[width=0.90\linewidth]{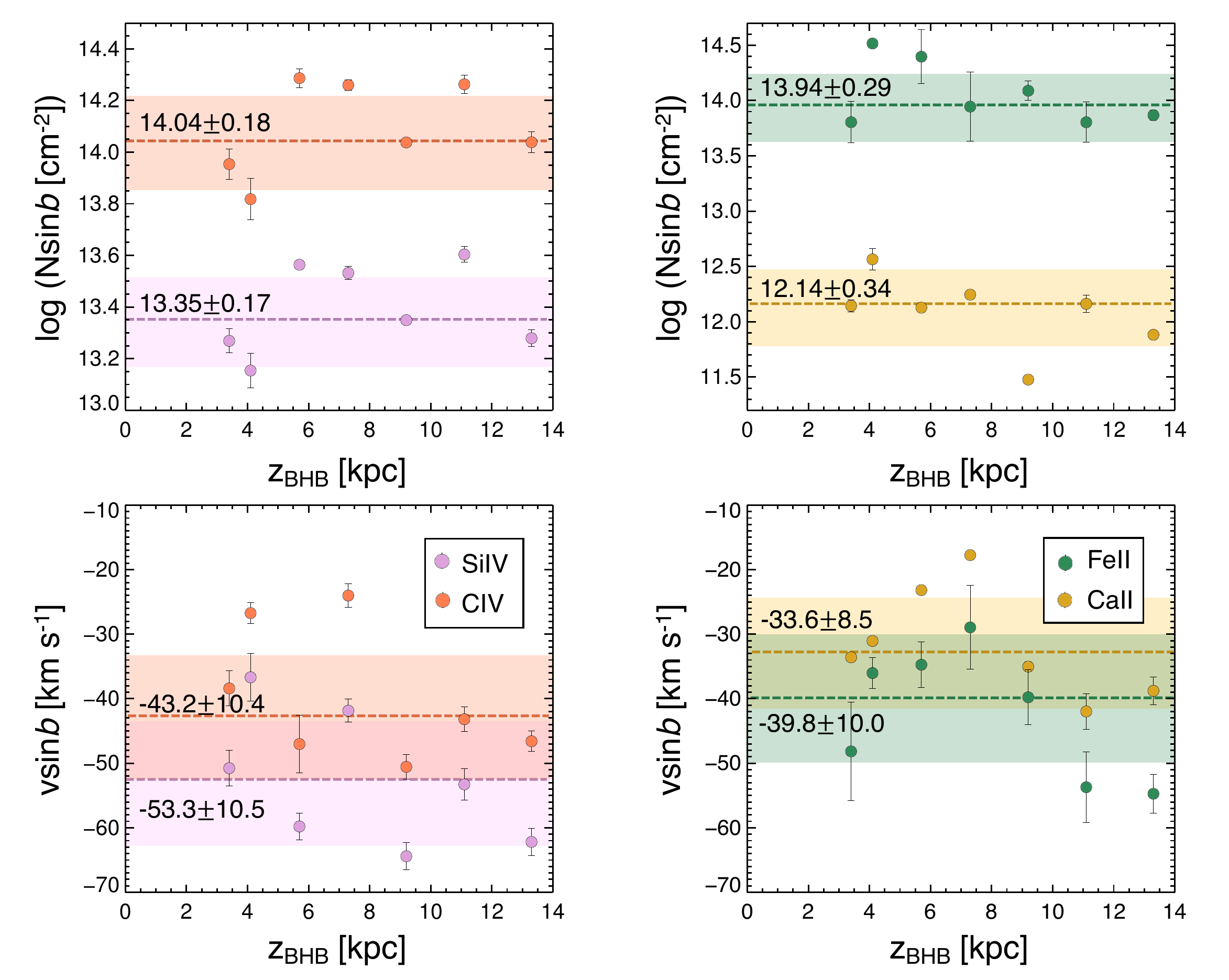}
\end{centering}
\caption{  Lack of trends with maximum height, given as z, from the MW-disk for derived properties of IV gas traced by both intermediate-ionization state species (SiIV, plum, CIV, coral, left-hand panels) and low ionization state species (FeII, green, CaII, gold,  right-hand panels). The top two panels examine vertical column density as a function of z, with the correction factor for Galactic latitude applied. The bottom panels show variations in the velocity centroids of the IV components, again corrected for Galactic latitude, $b$.  Median values, as dashed lines, with standard deviations, delineated by the shaded regions, of log Nsin$b$ and vsin$b$ are shown in the corresponding colors on each panel.  There are no significant trends with height for these quantities (regardless of the sin$b$ scaling), implying that the majority of the gas in absorption traced by these observations lies at z $\lesssim$ 3.4 kpc.    }
\label{fig:scaleheight}
\end{figure*}

The estimated distance and position of each BHB in our sample allows us to place an upper limit on the distance and height to the observed IV gas along the line of sight. Examining how the measured absorption properties (N, $v$, Doppler $b$) vary with both distance upper limits and height upper limits may provide insight into the three-dimensional distribution of the gas. For example, an increase in column density with $z$ would signify gas located at the full range of scale heights probed by our BHBs ( 3 kpc $<$ $z$ $<$ 14 kpc). However, we find no significant correlation between any of these quantities and either distance or height. We note, however,  that our data are not sensitive to variations in column density below 3.4 kpc. Below, we limit our discussion to the Galactic-latitude corrected ``vertical" projections (e.g. log Nsin$b$, vsin$b$) and height. 

The top panels of Figure \ref{fig:scaleheight} show that the vertical component of the IV gas column density in every ionization state, from CaII to CIV, remains roughly constant over the range of heights probed by our observations. Each panel shows a median of the value for log Nsin$b$ or vsin$b$ for each ion in the IV component only.  For the low ions, we show only CaII and FeII IV  absorption properties since these two ions have the best-constrained IV Voigt profile fits.  There is only small scatter of $\pm$ 0.2 dex for the intermediate ions (top left) and a more moderate scatter of $\pm$ 0.3 $-$ 0.4 dex for the low ions (top right). This narrow span of low and intermediate-ion column densities indicates that the majority of the gas traced by absorption lies at heights $\lesssim$ 3.4 kpc.  

 The lack of any discernible increase in column density with $z$ above 3.4 kpc is in agreement with the analysis of Savage and Wakker (2009\nocite{savage09}).  In their study of numerous halo star and QSO sightlines, they determined an exponential scale height of a given ion by comparing column densities projected along the z axis to a simple `plane-parallel' model of column densities with an exponential decline in the z direction. In this plane-parallel framework, they find that SiIV and CIV have similar exponential scale heights of 3.2 and 3.6 kpc, with typical errors on the order of $\pm$1 kpc.  Low ions tend to exhibit much smaller exponential scale heights $<$ 1 kpc. 

The median values of log Nsin$b$ of both SiIV and CIV along our BHB sightlines agree with the median log Nsin$b$ reported by Savage and Wakker (2009) for extragalactic lines of sight with \emph{b} $>$ 45$^{\circ}$. Their median values of log Nsin$b$ are 13.63$\pm$0.21 and 14.12$\pm$0.25,  for SiIV and CIV respectively, while our corresponding values of log Nsin$b$ are 13.35$\pm$0.17 and 14.04$\pm$0.18. We note that the apparent optical depth method column densities they report include any contribution from the MW itself, which is consistent with our median values of log Nsin$b$ being slightly lower than theirs. Adding back in the MW contribution to our values for log Nsin$b$, we get medians of 13.48$\pm$0.22 and 14.10$\pm$0.24, for SiIV and CIV respectively. This additional contribution brings our median vertical column densities into even better agreement with those of Savage and Wakker (2009). 

\begin{figure*}[t!]
\begin{centering}
\hspace{-0.4in}
\includegraphics[width=1.1\linewidth]{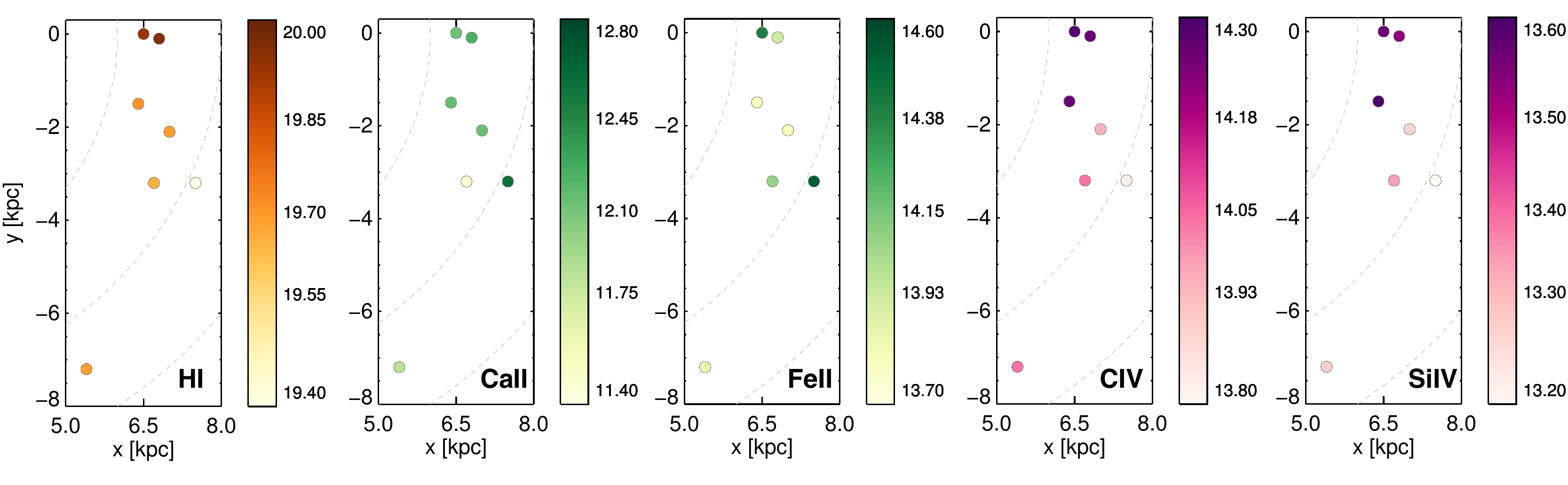}
\end{centering}
\caption{Maps in the XY plane of intermediate-velocity (IV) vertical ionic column densities (log Nsin$b$), including the EBHIS 21-cm log Nsin$b$,  at the position of each BHB. In this reference frame, the MW center is at the origin (0, 0) off the upper left-hand side of the Figure. The physical scale on each axis is given in kpc assuming a distance of 3 kpc for the IV gas.  The associated color bar on the right of each panel gives the column density range. The low ions (CaII, FeII) show a larger scatter in column densities and little coherence. In the intermediate ions (SiIV, CIV) and HI, the structure is more coherent, with a decrease in vertical column density of $\sim$0.05 dex per kpc from top to bottom.  For reference, gray dashed curves show circles of constant galactocentric radius centered on (x,y) $=$ (0,0) plotted every 2 kpc.  }
\label{fig:ionmaps}
\end{figure*}

 There is minor tension, at the $\sim$1$\sigma$ level, between the best-fit vertical velocity values for the intermediate (SiIV) and low ions (CaII) seen in the bottom two panels of Figure \ref{fig:scaleheight}. This tension increases when we consider IV HI-21cm emission radial velocities (ranging from  $-40$ $<$ v$_{\rm LSR}$ [km s$^{-1}$] $<$ $-15$). The IV gas traced by intermediate-ion absorption appears to be moving toward the disk of the Milky Way approximately 10 km s$^{-1}$ faster than the material traced by the low ion absorption. This offset could be due to systematic uncertainty introduced by our Voigt profile fitting assumptions not accounted for by the errors propagated during the iterative fitting procedure. It may also arise if we have underestimated the wavelength calibration uncertainty (but see Figure 2). While there may be physical implications of a velocity offset between different gas phases, we do not comment on it further given its low significance in our data.

\subsection{Column Density Variation in the XY plane}
\label{sec:columns}

Having established that our data show no significant trends with z, here we focus on a comparisons between column densities of the fitted intermediate velocity components as a function of their transverse separation from one another. Column densities of SiIV and CIV, with ionization potential energies of 45.1 eV and 64.5 eV, respectively, exhibit the strongest correlation with each other from sightline to sightline.  For log N$_{\rm SiIV}$ vs. log N$_{\rm CIV}$, the Spearman's Rho rank correlation coefficient (a value of 1 $=$ perfectly increasing monotonic correlation) is 0.86 and the linear Pearson correlation coefficient  (a value of 1 $=$ perfectly increasing linear correlation) is 0.97. Figure \ref{fig:ionmaps} clearly shows that the seven IV column density measurements of SiIV and CIV are strongly correlated with each other. The low ions (e.g. CaII, FeII) with ionization potential energies $<$ 20 eV exhibit weak to moderate correlations with each other (Spearman's Rho $=$ 0.2 - 0.3; Pearson's coefficient $=$ 0.3 - 0.5), the larger scatter likely due to the large uncertainties in the fitted column densities of the weak IV component.

Next, we examine whether the column densities of low ionization state gas traced by SiII, FeII, and CaII correlates with the columns of intermediate-ionization state gas traced by SiIV and CIV. The Spearman's Rho rank correlation coefficient between log N$_{\rm lowion}$ and log N$_{\rm highion}$ ranges between -0.1 and 0.1 for the various species, strongly indicating no monotonic correlation of any kind between the species of different ionization states. As we will explore in the Discussion Section, this result is consistent with pictures put forth by many previous studies of the Milky Way halo and Magellanic Stream \citep[][]{putman03, fox04, richter17} in which larger envelopes of ionized gas contain patchy clumps of cooler, more neutral gas. 

\begin{figure*}[t!]
\begin{centering}
\hspace{-0.2in}
\includegraphics[width=1.0\linewidth]{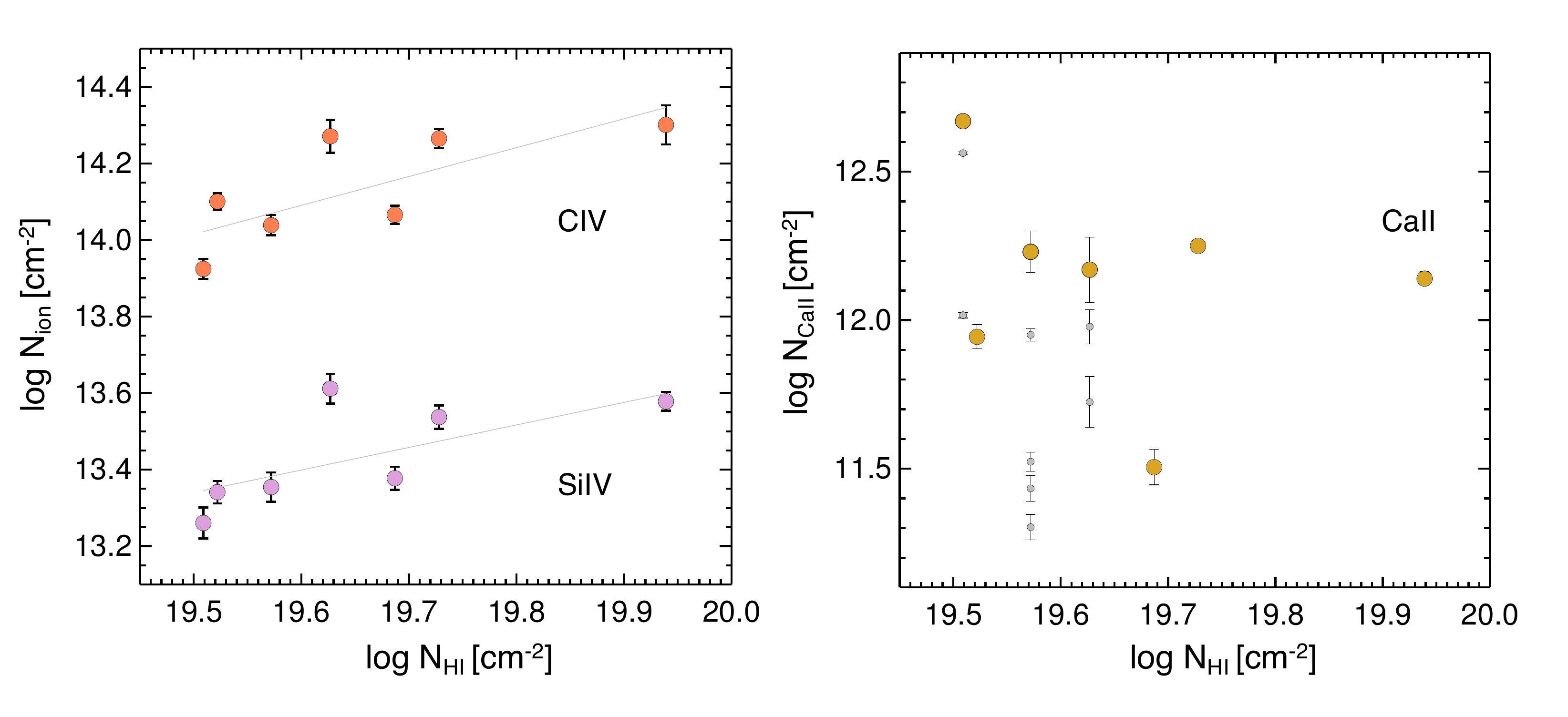}
\end{centering}
\caption{Left: Correlation between column densities of intermediate ions CIV (coral) and SiIV (plum) with the EBHIS 21-cm log N$_{\rm HI}$ in the intermediate velocity (IV) gas components.  The linear least squares fits are showns as  grey lines, giving positive slopes of 0.59$\pm$0.076 (SiIV) and 0.75$\pm$0.096 (CIV). Right: We show the column densities of all individual CaII IV components (in gray) and the sum of the CaII IV component column densities along BHB lines of sight (in gold). There is no correlation between CaII IV component column densities and the IV EBHIS 21-cm log N$_{\rm HI}$. Although we do not show it here, there is a similar lack of a trend and large scatter in other low ion column densities measured in the COS data (e.g. FeII, SiII), albeit with considerably larger errors in the fitted IV component columns. }
\label{fig:nionnhi}
\end{figure*}

Finally, we search for trends between ionic column densities and the IV log N$_{\rm HI}$ in order to assess whether the gas seen in absorption toward the BHBs is tracing the same material as that seen in emission in Figure \ref{fig:HIImage}. First, it is important to recall that the HI column densities based on the EBHIS 21-cm emission data are averaged over a much larger angular scale than the absorption-line column densities.  Furthermore, the precise value of N$_{\rm HI}$ associated with N$_{\rm ion}$ is impossible to determine using the 21-cm data alone.  Any dense clumps or `cloudlets'  along each line of sight with angular sizes $<$ 10.8$\arcmin$ would be washed out by the larger EBHIS beam \citep{wakker01b, lockman02, saul12}. Any differential depletion would add additional scatter. For these reasons, it is not surprising that gas column densities of the low ions traced by CaII, FeII, and SiII in Keck and COS data do not correlate at all with N$_{\rm HI}$. We show this lack of correlation on the right-hand panel of Figure \ref{fig:nionnhi} for CaII, which offers the best  constraints on the absorption properties of the low ions. 

There is, however, a moderate to strong correlation between {\emph{both}} SiIV and CIV and HI, seen on the left-hand panel of Figure \ref{fig:nionnhi}. For both intermediate ions, the linear least squares fit is shown as a grey line, indicating positive slopes of 0.59$\pm$0.076 (SiIV) and 0.75$\pm$0.096 (CIV). The Spearman's rho rank coefficient (such that a value of 1 $=$ a perfectly monotonic positive correlation) in each case indicates a moderately strong correlation, with values of 0.78 and 0.75 for SiIV and CIV, respectively. What this likely means is that the HI column densities in the 10.8$\arcmin$ beam on average are tracing the larger-scale, diffuse structures rather than the smaller-scale, dense `cloudlets' probed by the low ions.  

\begin{figure*}[t!]
\begin{centering}
\hspace{-0.2in}
\includegraphics[width=1.0\linewidth]{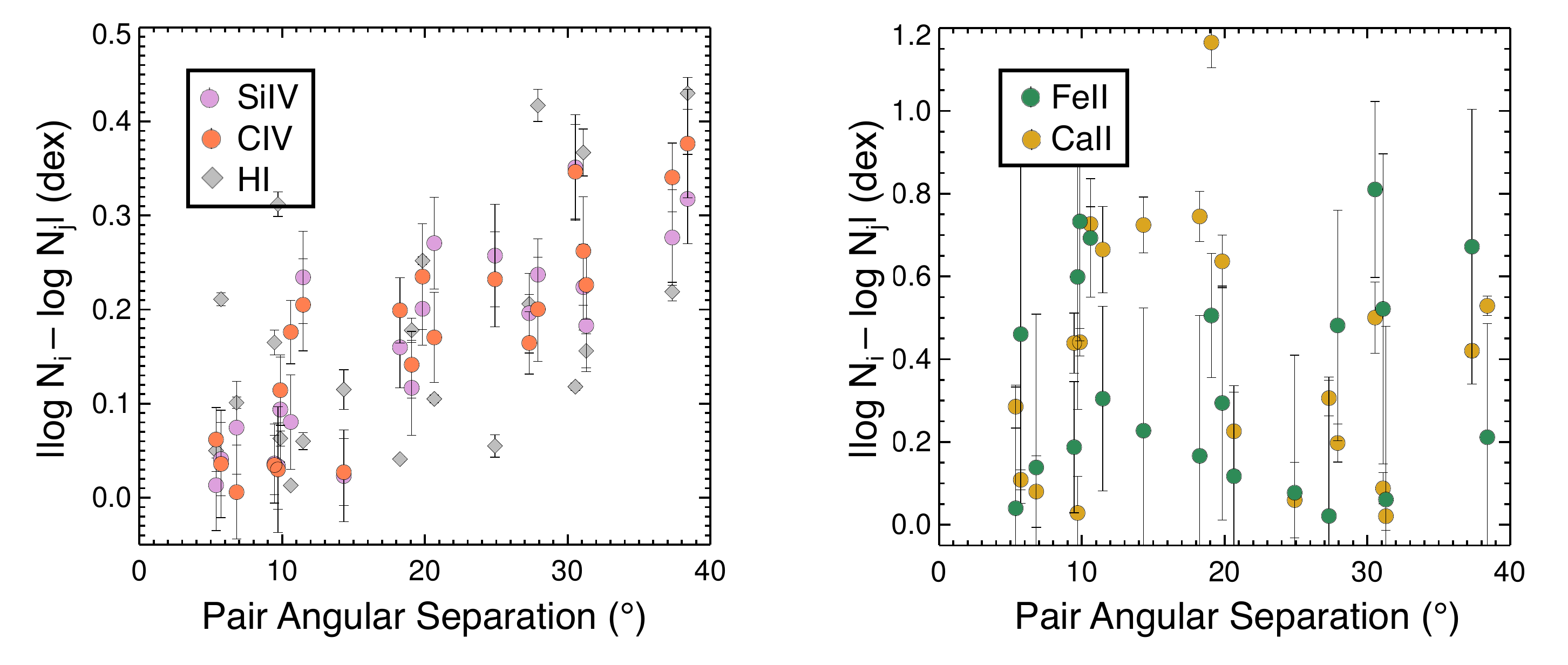}
\end{centering}
\caption{ Pairwise ionic column density differences for the IV components along all 7 sightlines (21 unique pairs) vs.\ their angular separation on the sky in degrees. At a distance of 3.0 kpc, 10 degrees corresponds to $\sim$0.5 kpc.  There is a highly significant monotonic, increasing correlation between pair separation and column density differences in the intermediate ions (SiIV in plum, CIV in coral; Spearman's rho $\approx$ 0.82 for both). HI 21-cm emission (in gray diamonds; Spearman's rho $\approx$ 0.46)  shows only a weak-to-moderate correlation with much larger scatter. In contrast, on the right panel, there is no correlation between low-ion column density differences between pairs and their angular separation on the sky, and much larger  column density differences overall (shown: FeII in green, CaII in gold; Spearman's rho $\lesssim$ 0.05).     }
\label{fig:deltacol}
\end{figure*}

Directly addressing the angular scale of the absorbing and emitting structures, Figure \ref{fig:deltacol} shows the difference in column density between each unique IV component `pair' for the 7 lines of sight (21 pairs total) versus the pair angular separation. The CIV and SiIV are tracing a coherent, large-scale structure, as evidenced by the strong correlation between $\Delta$logN and the separation over 40$^\circ$ degrees. Both intermediate-ion correlations give a Spearman's rho coefficient of 0.82, with a two-sided significance of its deviation from 0 to be $<$ 0.000004.  The scatter in this correlation is only $\sim$0.15 dex, and within 10$^\circ$  the median $\Delta$logN is only 0.05 dex, indicating a relatively smooth structure. Furthermore, the IV HI 21-cm column densities (shown for reference as grey diamonds)  seem to follow a similar trend, however with a slightly larger scatter. At a distance of 3 kpc, roughly the upper limit we can set with our data, 10$^\circ$ $\approx$ 0.5 kpc. At the lower distance limit to the IV arch of 0.9 kpc \citep{wakker01a}, the structure of which the BHB sightlines skirt the edge, 10$^\circ$ $\approx$ 0.15 kpc. 

On the right-hand panel of Figure \ref{fig:deltacol}, we show the same pairwise $\Delta$logN versus angular separation for two low ionization state transitions, CaII (gold) and FeII (green).  While the CaII IV component column densities are well-determined from the Keck HIRES data, the log N$_{\rm FeII}$ of the IV component has a much larger error owing to the many assumptions and limits imposed on the lower resolution COS spectrum. Nonetheless, we can rule out any correlation between $\Delta$logN$_{\rm lowion}$ and angular separation with high confidence. Furthermore, the scatter of the data over the 40$^\circ$ angular scales is $>$ 0.4 dex. This large scatter and lack of a correlation are consistent with patchy, small-scale clumps of material. Considering the monotonic increase with small scatter to 40$^\circ$ of the intermediate ions, and the large scatter and lack of trend seen in the low ions (right panel of  Figure \ref{fig:deltacol}) our data directly support the physical picture of large ionized envelopes or streams containing cooler condensations or clumps. 
 
\begin{figure}[t!]
\begin{centering}
\includegraphics[width=1.0\linewidth]{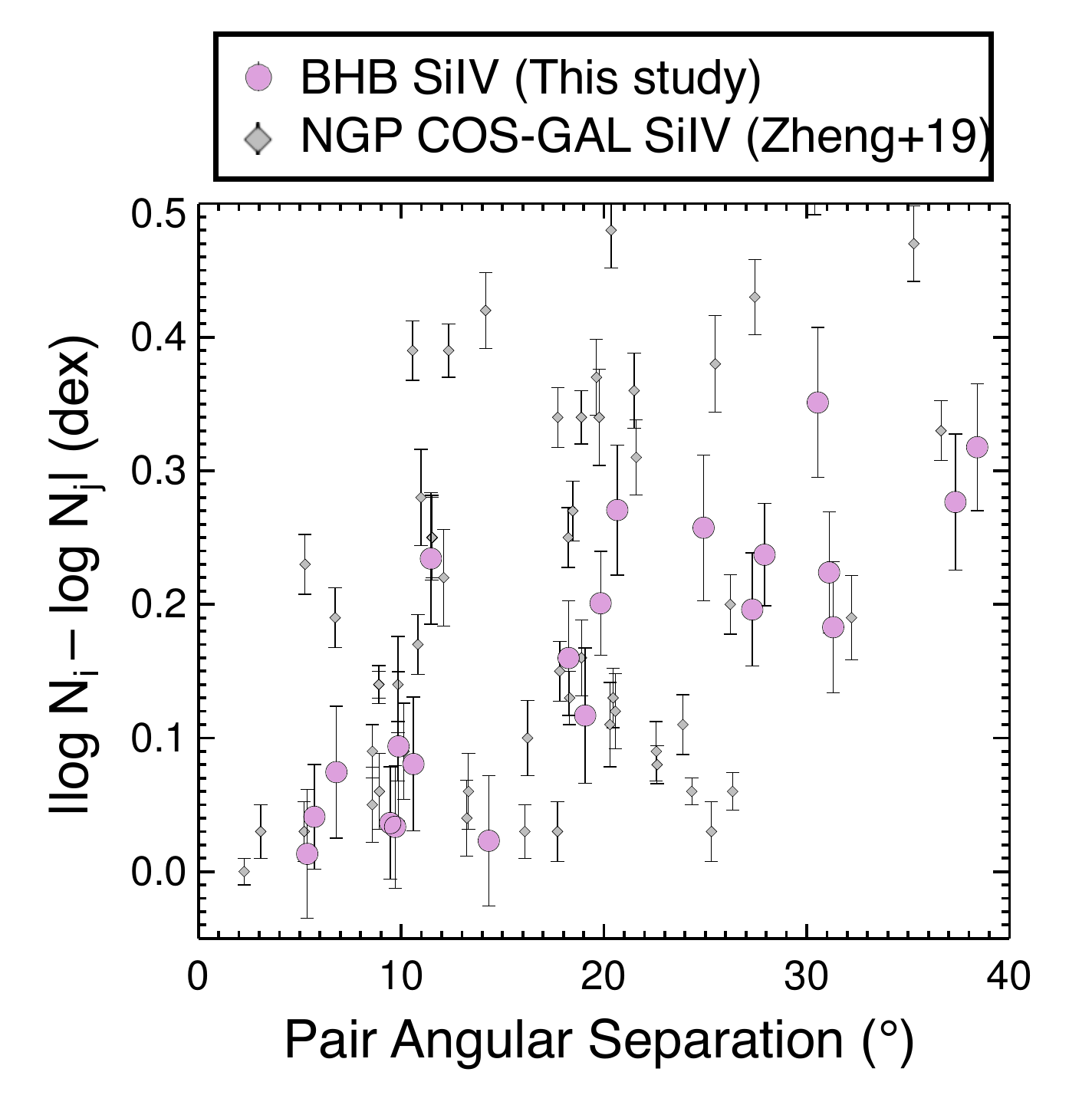}
\end{centering}
\caption{ Pairwise SiIV column density differences for the IV components along all 7 sightlines (plum circles; 21 unique pairs) vs.\ their angular separation on the sky in degrees, compared to a sample of 11 sightlines from the COS-GAL database (gray diamonds; \citealt{zheng18}) selected to be in the same region of the Northern Galactic sky as the BHBs. The scatter in the COS-GAL QSO sightline database is larger for reasons outlined in Section \ref{sec:columns}, but there remains a weak-to-moderate increasing correlation between pair column density difference and angular separation, with a Spearman's rho correlation coefficient of 0.35.     }
\label{fig:deltacollargearea}
\end{figure}

In principle, additional archival UV absorption-line data could support this picture of $\sim$ kpc-scale ionized structures with even larger area coverage. As an initial step, we harvested SiIV column densities from the COS-GAL QSO sightline database \citep{zheng18} to explore $\Delta$logN variations over a similar area in the Northern Galactic sky to that of our BHBs. In total, we selected 11 COS-GAL sightlines against which to compare our BHB SiIV IV-component column densities. The results can be seen in Figure \ref{fig:deltacollargearea}, where the gray points track the pairwise SiIV column density differences for the extragalactic COS-GAL sightlines. This archival sample shows only a  weak-to-moderate increasing correlation between pair column density difference and angular separation, with a Spearman's rho correlation coefficient of 0.35, and a two-sided significance of its deviation from 0 to be $\approx$0.04.

The most likely reason for the weaker correlation in Figure \ref{fig:deltacollargearea} is that there is a contribution to the absorption along the QSO sightlines from the more extended, ionized CGM up to much larger heights, $\sim$200 kpc \citep{zheng18}. Each QSO sightline probes the full MW halo virial radius, and likely pierces multiple kpc-scale structures in SiIV. The effect would be to wash out the individual structure coherence seen for our stellar sightlines. Furthermore, the apparent optical depth method column densities reported by COS-GAL include a MW component, whereas we explicitly do not. The MW component's contribution to the total column density is minimal for SiIV and CIV but still exists at the 20\% level, and likely adds additional scatter to the relation. Thus, we consider Figure \ref{fig:deltacollargearea} to offer tentative support for our physical picture (explored in detail in the Discussion). Further support would come from analyzing all stellar sightlines with distance constraints probed by COS and STIS, including a full Voigt profile analysis similar to that which we present in this work. Such an analysis over large angular area, while beyond the scope of our study, would potentially be sensitive to the {\emph{maximum}} size scale of these large, ionized structures. 

\section{Ionization Models}

As intermediate ionization species, SiIV and CIV may be a part of the warm ionized medium (WIM, 10$^4$ $-$ 10$^5$ K) or may lie in a separate, tenuous ``transitional" phase at $\sim$10$^5$K. The former would indicate photoionization, while the latter would point toward collisional ionization, or gas cooling out of the ambient hot halo,  as the dominant process giving rise to the SiIV and CIV. In this section, we examine the constraints imposed by our observations for both cases. Ultimately, this analysis drives at the origin of the material and possibly at the nature of its surroundings, informing our understanding of relevant Galactic fountain astrophysics.

\subsection{Constraints from Photoionization Models}
\label{sec:ionmodel}

Photoionization models, such as those constructed with the spectral synthesis code CLOUDY \citep{ferland17},  perform detailed radiative transfer through a parcel of gas characterized by some neutral column density, N$_{\rm HI}$,  metallicity, [M/H], and ionization parameter, U (U $\equiv$ (n$_{\rm H}$ $\times$ c)/$\Phi_{\rm tot}$; where  $\Phi_{\rm tot}$ is the integrated ionizing photon flux generated by the assumed photoionizing spectrum). One must make several assumptions to run such a CLOUDY model: 1) thermal and ionization equilibrium; 2) an ionizing background radiation field; and 3) a density structure for the parcel of gas, often approximated as a uniform density slab (but see Stern et al. 2016\nocite{stern16}). Another common assumption of these models is relative solar abundances of the elements, although these element ratios can be allowed to vary to account for depletion onto dust \citep{jenkins09}. Comparing the outputs of a grid of ionization models with the data (e.g. systematically varying N$_{\rm HI}$, [M/H], and/or U) ultimately allows one to find the model or set of models that are consistent with the constraints set by the ionic column densities determined from the observations.

We construct such a grid of CLOUDY models to constrain the physical state of the IV gas probed by our observations of SiIV and CIV {\emph{only}}, assuming they are predominantly photoionized (we address collisional ionization in and out of equilibrium in \S \ref{sec:collide}). We assume solar ratios of the elements, and exclude the low ions from this analysis. We are justified in doing so as our previous analysis has shown that they are almost certainly not co-spatial with the intermediate ionization state material. We note that if both the low ions and the intermediate ions are part of the photoionized WIM, then the WIM material itself cannot be characterized by a single density ``slab." However, our CLOUDY models of the single-phase intermediate ionization IV gas assume a uniform density layer irradiated by the extragalactic UV background and escaping radiation from the Milky Way disk, as presented by \cite{fox05}. These models represent a simplified picture that nonetheless may prove informative. 

$\Phi_{\rm tot}$ (the number of ionizing photons emitted per second by the assumed radiation field) at the Milky Way disk-halo interface is highly uncertain and depends strongly on: (1) the absorbing feature's distance from the Milky Way and (2) an assumed escape fraction of the ionizing photons from the disk. In practice, adjusting $\Phi_{\rm tot}$  affects primarily the calculated hydrogen number densities, n$_{\rm H}$ = $\Phi_{\rm tot}$/ (U $\times$ c). To address this uncertainty, we ran several CLOUDY models, assuming  the Fox et al. (2005) ionizing spectrum at different heights above the disk. For reference, we find that at 10 kpc, 2 kpc, and 0 kpc above the MW disk, $\Phi_{\rm tot}$ $=$ 340000 s$^{-1}$, 5500000 s$^{-1}$ and  37000000 s$^{-1}$, respectively. We note that the ionization parameters and ionized fractions of hydrogen and metals are consistent among these models due to identical {\emph{spectral shapes}}.  However, it is apparent that such a range in disk heights produces ionizing photon fluxes, and thus gas volume densities, that vary over two orders of magnitude. Specifically, the assumed hydrogen number densities shift upward by two orders of magnitude with a corresponding increase in $\Phi_{\rm tot}$.  Thus, constraining the volume density of this IV material is not possible without additional constraints on key parameters (namely, N$_{\rm H}$ and [M/H], aka metallicity). In the following analysis, we report metallicities relative to solar abundances, X[M/H]/[M/H]$_{\odot}$ using the compact notation [M/H]$=$ X where X is some fraction of solar.


\begin{figure*}[t!]
\begin{centering}
\hspace{-0.2in}
\includegraphics[width=1.0\linewidth]{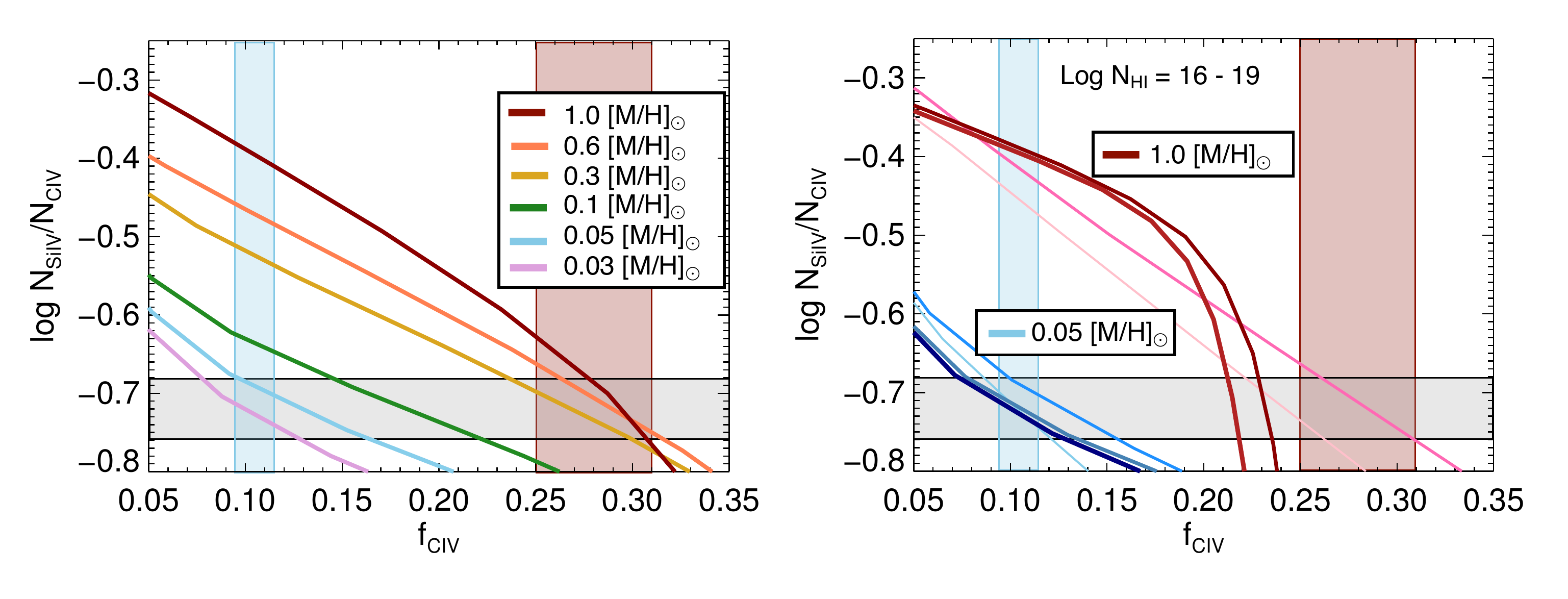}
\end{centering}
\vspace{-0.1in}
\caption{  The fraction of carbon seen as CIV, f$_{\rm CIV}$,  versus the column density ratio log N$_{\rm SiIV}$/N$_{\rm CIV}$ for six CLOUDY photoionization models of varying metallicity [M/H]  $=$ 0.03 $-$ 1.0 (left) and varying log N$_{\rm HI}$ from 16 $-$ 19 in 1 dex steps (right). The range of observed log N$_{\rm SiIV}$/N$_{\rm CIV}$ is shown in the gray horizontal regions on each panel, with well-constrained values ranging from $\approx$ $-0.76$ to $-0.68$.   On the right-hand panel, we show increasing log N$_{\rm HI}$ [cm$^{-2}$] with line thickness and darkness from 16 - 19. We show these sets of curves at two values of metallicity, [M/H] $=$ 0.05 (blue curves) and  [M/H] $=$ 1.0  (brownish-red curves). We highlight two potential ``solutions" for gas-phase metallicity, f$_{\rm CIV}$ $\approx$ 11\% in vertical bands of light blue, and f$_{\rm CIV}$ $\approx$ 25\% $-$ 31\% in brownish-red. We discuss these two divergent solutions in \S \ref{sec:ionmodel}.   }
\label{fig:ionmodel}
\end{figure*}

Under the above assumptions, the column density ratio  log N$_{\rm SiIV}$/N$_{\rm CIV}$ is a sensitive probe of the fraction of total carbon seen as CIV $\equiv$ f$_{\rm CIV}$. This dependence is shown in Figure \ref{fig:ionmodel}. The median log N$_{\rm SiIV}$/N$_{\rm CIV}$  = $-0.69\pm0.04$, and shows little variation from sightline to sightline. The precise value of f$_{\rm CIV}$ inferred by the measured range of $-0.76$ $<$ log N$_{\rm SiIV}$/N$_{\rm CIV}$ $<$ $-0.68$ depends strongly on the assumed metallicity of the gas. This metallicity dependence is a result of the shape of the cooling curve, and the strong dependence of the high and intermediate ionic fractions on the metal line cooling. The left-hand panel shows these models are very sensitive to the gas-phase metallicity. We  chose an N$_{\rm HI}$ $=$ 17.5 at which to plot these CLOUDY models on the left, and note that the models are only sensitive to the choice of N$_{\rm HI}$ at  [M/H] $\gtrsim$ 1.0 (at metallicities $\lesssim$ solar, there is no sharp turnover for optically thick models,  and the spread in f$_{\rm CIV}$  looks similar to the blue curves on the left panel). On the right-hand panel, we see that these curves shift slightly with variations in assumed N$_{\rm HI}$ over a range of  16.0 $<$ log N$_{\rm HI}$ $<$ 19.0, with greater differences between optically thin and thick gas at high metallicity.    The fraction of Carbon in CIV does not depend at all on the assumed $\Phi_{\rm tot}$ (although it does depend on the slope of the assumed ionizing spectrum). For the range of log N$_{\rm SiIV}$/N$_{\rm CIV}$ in the observed IV gas along our BHB sightlines, these models indicate that anywhere between 7\% and 32\% of the carbon is in CIV. Thus, for our median log N$_{\rm CIV}$sin$b$ of 14.04, this implies  14.5 $<$ log N$_{\rm C, total}$sin$b$ $<$ 15.2. In principal, if we had some way of independently constraining either the metallicity of the material or the total hydrogen column density along the line of sight, N$_{\rm H}$, we would fully constrain the metal content of the IV gas along our sightlines.

\subsubsection{Independent Constraints on [M/H] and N$_{\rm H}$}
 
 Addressing the metallicity of the gas,  Wakker et al. (2001, 2001b\nocite{wakker01a, wakker01b})  present a thorough and exhaustive spectroscopic analysis to determine the metallicity of the nearby IV arch and other IVC and HVC halo structures. They use reliable values of N$_{\rm HI}$ and measurements of NI, OI, and SII along four lines of sight, all low ions that are presumed to trace predominantly neutral material. \cite{wakker01a} finds that the neutral phase of the IV arch has a metallicity of 0.8 $-$ 1.2 solar. In general, IVCs present metallicities typically between 0.5 $-$ 1.5 solar, indicating an origin in the disk of the Milky Way \citep{richter17}. 
 
 Assuming the IV neutral and ionized gas phases are ``well-mixed," they would have the same metallicity. There are at least a few lines of evidence to suggest this is not an unreasonable approximation \citep[e.g.][]{richter17}.   Adopting solar metallicity for the intermediate ionization state material traced by our observations puts us on the brownish-red curve on the left-hand panel in Figure \ref{fig:ionmodel}.  This curve then leads us to adopt the maximum value of f$_{\rm CIV}$, $\sim$30\%, where the constraints of our measurements on  log N$_{\rm SiIV}$/N$_{\rm CIV}$ indicated by the horizontal gray shaded region intersect with the curves shown by the vertical rectangular brownish-red-shaded region. Thus, we may adopt the lower bound on the total carbon content of the IV gas of  log N$_{\rm C, total}$sin$b$ $\approx$ 14.5. For self-consistency with the assumed solar metallicity, considering the solar carbon abundance of 12 $+$ log (C/H) $=$ 8.43, this estimate implies an N$_{\rm H}$sin$b$ of $\approx$ 18.1. 

However, we may additionally constrain N$_{\rm H}$sin$b$ using an entirely different method. To do so,  we turn to pulsar dispersion measure estimates for the Milky Way halo \citep[e.g.][]{ne2001, howk06, gaensler08}. As the light from a pulsar located in the Milky Way halo interacts with free elections along its line of sight, pulses at lower frequencies experience a delay relative to the higher frequency pulses \citep[e.g.][]{ferriere01}. Sensitive pulsar timing experiments toward thousands of pulsars in the Milky Way halo allow constraints on the dispersion measure (DM) of radio pulsars that translates directly into a free-electron column, N$_{\rm e}$ \citep[e.g.][]{taylorcordes93}.  Several works have combined these measurements into a physical model of the Milky Way warm ionized medium (WIM) by assuming some large-scale structure (e.g. exponential disk).  These models can then be used to estimate N$_{\rm e}$ at any given set of galactic coordinates for any assumed height \citep[e.g.][]{ne2001,bmm06, gaensler08,ymw16}. 

We use the model of \cite{gaensler08}, with the updates from Savage \& Wakker (2009)\nocite{savage09} that account for inhomogeneities in the gas,  to estimate N$_{\rm e}$sin$b$ for high-latitude lines of sight in the region of the halo probed by our BHB sightlines. Integrating the N$_e$ of the model to the maximum height of our BHB sample results in log N$_{\rm e}$sin$b$ $=$ 19.83 cm$^{-2}$. This value is similar to that of \cite{howk06}, who use a value of  log N$_{\rm e}$ toward their high-latitude pulsars of 19.91  cm$^{-2}$, although these lie at distances of $\sim$10 kpc. Accounting for the distance constraints on our IV gas clouds, determined to predominantly lie at z $<$ 3.4 kpc, we reduce our adopted value of  log N$_{\rm e}$sin$b$ to 19.79 (see Equation 6 of \citealt{gaensler08}, with z = 3400 pc, H$_{n}$ = 1410 pc and n$_{0}$ = 0.0155 cm$^{-3}$ as in \citealt{savage09}; Here, n$_{0}$ is the density of gas at the mid-plane, and H$_{n}$ is the corresponding scale-height). 

Additional assumptions, such as the fraction of free electrons due to helium, and the fraction of free electrons due to hot, coronal gas, allow us to convert  N$_{\rm e}$ to N$_{\rm H}$. For these additional assumptions, we rely on the pioneering work of \cite{howk06}, who used sightlines toward globular clusters containing radio pulsars to independently constrain N$_{\rm H}$ and derive accurate gas-phase abundances. We refer readers to that paper for details on the method we follow, and the limits we adopt.  For the helium correction, \cite{howk06} adopt a minimal case such that the fraction of doubly-ionized helium, x(He$^{+2}$) $=$ 0, and $\eta_{\rm WIM, min}$, a numerical correction factor to the conversion between N$_{\rm e}$ and N$_{\rm H}$, $=$ 0.98. The maximum correction factor for helium, $\eta_{\rm WIM, max}$ is assumed to be 0.81. Following a two component prescription for the hot ionized medium (HIM) contribution presented in Equation 3 of \cite{howk06}, we assume (N$_{\rm e}$sin$b$)$_{\rm HIM}$ $=$ 19.10. Thus, our minimum and maximum values of log N$_{\rm HII, WIM}$sin$b$ are 19.59 and 19.67, respectively. 

Converting  N$_{\rm HII}$sin$b$ to  N$_{\rm H}$sin$b$ requires knowing the neutral fraction, which requires we return to our ionization model. We note that for the full range of models shown in Figure \ref{fig:ionmodel}, over the allowed values of log (N$_{\rm SiIV}$/N$_{\rm CIV}$),  the neutral gas fraction ranges from $\sim$0.1\% $-$ 0.0001\%. Thus, we assume that  N$_{\rm HII}$sin$b$ $\approx$  N$_{\rm H}$sin$b$ for the intermediate ionization state material. Immediately we can see there is a mismatch with the solar metallicity model discussed earlier in this section, on the order of 1.5 dex (log N$_{\rm H}$sin$b$ of $\approx$ 18.1 vs 19.6)! Another way to examine this discrepancy is to find the set of self-consistent models such that the derived abundance from N$_{\rm C}$ / N$_{\rm H}$ is equal to the value one would derive from the [M/H] curve in Figure \ref{fig:ionmodel} and the associated f$_{\rm CIV}$. The only self-consistent models in this framework lie at [M/H] $\approx$ 0.05. Tuning the assumptions within the allowed ranges does not bring the two models into agreement. For example, adopting a metallicity at the low end of the neutral IVC metallicity range, 0.5 solar, and choosing a maximal value of (N$_{\rm e}$sin$b$)$_{\rm HIM}$ $=$ 19.22 (Howk et al. 2006), the values continue to be highly discrepant, on the order of $\sim$ 1 dex. 

 We note that these limits are robust to uncertainties in the photoionization modeling barring a very significant alternation to the assumed slope of the ionizing spectrum from 1 - 5 Ryd.  One way to resolve the discrepancy with ionization modeling is to arbitrarily assign SiIV and CIV to distinct phases themselves. Doing so would not be justified by the data, especially given the extent to which SiIV and CIV column densities, velocity centroids, and line-widths are well-matched along every single line of sight in our sample. Another possibility, which we explore in detail in \S \ref{sec:collide}, is that SiIV and CIV represent neither WIM nor HIM material, but rather are the manifestations of a tenuous, transitional-temperature, T$\sim$10$^{5}$ K plasma cooling out of the ambient $\sim$10$^{6}$ K coronal medium. 

 In summary, the two unacceptable photoionization model options presented above are that either the intermediate ionization state IV gas is metal poor and not well-mixed with the neutral IV gas or that (N$_{\rm e}$sin$b$)$_{\rm HIM}$ $\approx$ (N$_{\rm e}$sin$b$)$_{\rm WIM}$. That is, for the models to converge on a metal-rich solution such that [M/H]$_{\rm IV}$ $\approx$ 0.5 or above, the HIM or some other ionized gas phase would have to contribute $\gtrsim$ 95\% of the free electrons along the lines of sight to pulsars. Such a large value of log N$_{\rm HII, HIM}$ on the order of 19.7 is inconsistent with observed constraints on n$_{\rm H, HIM}$ $\lesssim$ 8 $\times$ 10$^{-4}$  cm$^{-3}$ \citep[][]{bregman07, grcputman09}.  Inefficient mixing of gas phases has been invoked in the context of Galactic winds \citep{oppenheimer12}, but is not a particularly elegant solution in a Galactic fountain scenario in which this material originated in the disk of the Milky Way. 
 
\subsubsection{A Self-Consistent Photoionization Solution Requires a Si-Depleted Ionized Medium}
Can we resolve this tension in the models, or must we reject photoionization of SiIV and CIV altogether? There is a solution we have not yet considered. That is, {\emph{the photoionization models are perfectly self-consistent if we assume that there is a substantial fraction of Silicon depleted onto dust in the warm-ionized disk-halo interface of the Milky Way. }}Turning again to Figure \ref{fig:ionmodel}, we see that the orange and brownish-red curves (metallicities of 0.5 solar $-$ 1.0 solar) are brought into consistency with the pulsar dispersion constraints (light blue shaded region) when we allow for Silicon depletion relative to Carbon between $-$0.25 and $-$0.35 dex. That is, the undepleted ratio of 
log N$_{\rm SiIV}$/N$_{\rm CIV}$ should be closer to $\approx -0.4$ if the gas is metal-enriched at a level of 0.5 solar $-$ 1.0 solar. 

Addressing the photoionization model's self-consistency, we note that the length scale implied by this fiducial model ($\ell = $ N$_{\rm H}$/n$_{\rm H}$) is: 
\begin{equation}
    \ell \approx 3.2 \rm ~kpc \left(\frac{N_{\rm H}}{10^{19.7} cm^{-2}}\right) \left(\frac{10^{-2.3} cm^{-3}}{n_{\rm H}} \right).
\end{equation} 
This value of n$_{\rm H}$ $=$ 0.0055 cm$^{-3}$ occurs at $\Phi_{\rm tot}$ $=$ 5500000 s$^{-1}$ (the value from a Fox et al. 2005 spectrum 2 kpc from the disk) precisely at an f$_{\rm CIV}$ of 11\%. While there is a slight mismatch (2 kpc vs. 3.2 kpc), the length scales are consistent enough within the uncertainty of $\Phi_{\rm tot}$ not to raise any red flags. Additionally, a $\approx$ 3 kpc length scale is consistent with the scale heights of SiIV and CIV given by \cite{savage09}. Finally, we do note that this gas density is more than an order of magnitude lower than the volume density of the warm, ionized interstellar medium (WIM) of the Milky Way disk inferred from measurements of singly and doubly ionized species \citep{reynolds93, sembach00}. If IV SiIV and CIV are an extension of the Milky Way's photoionized WIM, they would represent a lower density extension \citep[e.g.][]{stern16}.

\subsection{The Case in Which SiIV and CIV are Collisionally Ionized in Transition Temperature Gas}
\label{sec:collide}

\begin{figure}[t!]
\begin{centering}
\hspace{-0.3in}
\includegraphics[width=1.1\linewidth]{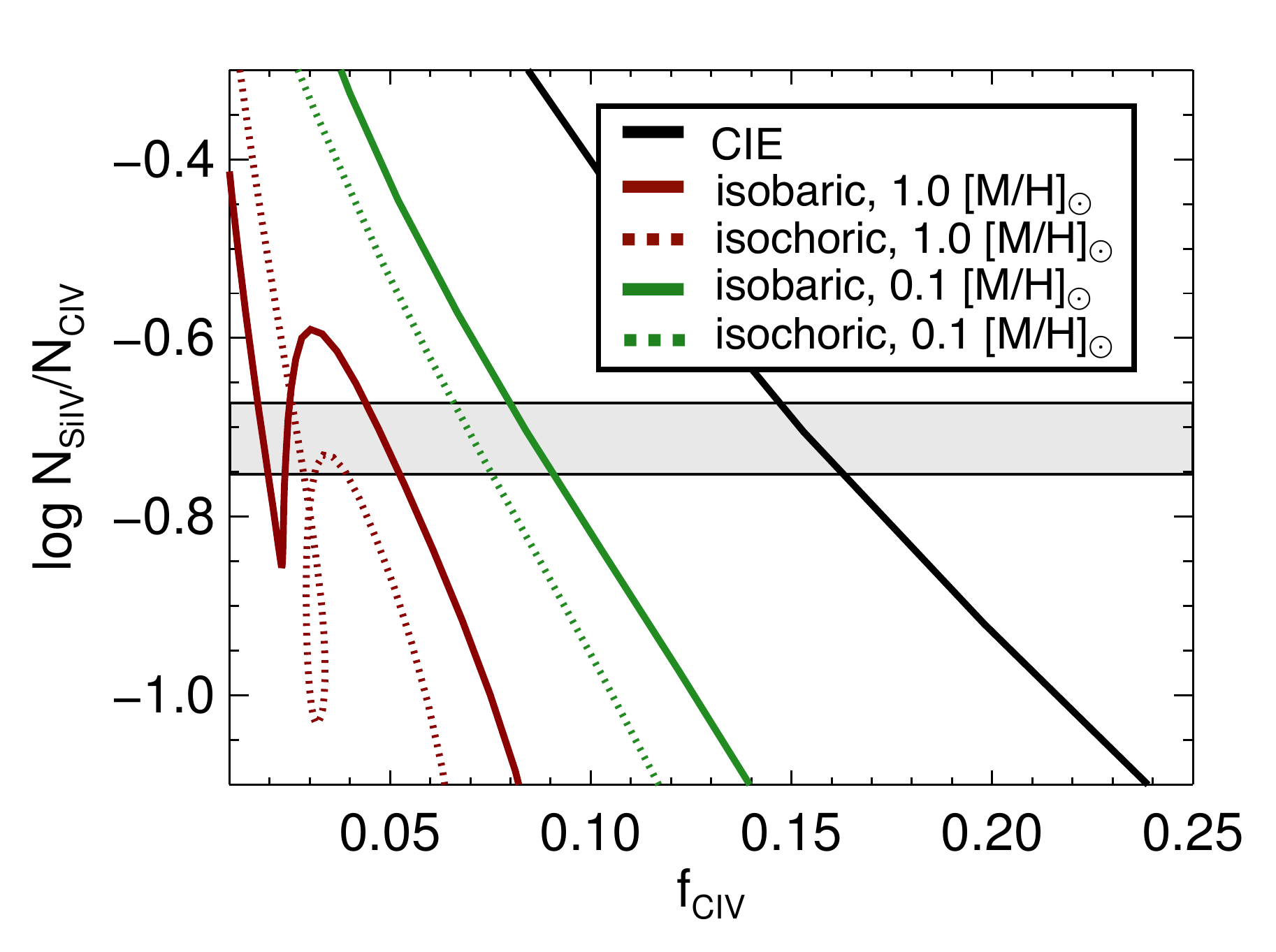}
\end{centering}
\vspace{-0.1in}
\caption{  The fraction of carbon seen as CIV, f$_{\rm CIV}$,  versus the column density ratio log N$_{\rm SiIV}$/N$_{\rm CIV}$ for several models of collisional ionization based on \cite{gnat07}.  Collisional ionization equilibrium (CIE) produces a trend shown as the black line, and is independent of density and metallicity. Out of equilibrium (brownish-red and green curves) the predicted column densities and ionization fractions depend on the metallicity of the gas and whether its cooling is isobaric or isochoric. The range of our measured  N$_{\rm SiIV}$/N$_{\rm CIV}$  is shown by the horizontal gray rectangle.   }
\label{fig:collisionionfrac}
\end{figure}

Collisional ionization is the dominant process in low-density plasma in the absence of a significant ionizing radiation field. In equilibrium (CIE), the ion fractions produced by this process are a function of only gas temperature, and are determined by detailed balance between electron impact collisional ionization, electron-ion recombination, and charge transfer reactions \citep{gnat07}. Outside of equilibrium, the ionization state of the plasma depends on whether the gas is cooling at a constant density (isochoric) or at a constant pressure (isobaric), and also on the gas-phase metallicity \citep[e.g.][]{wiersuma09, gnat09}.  Furthermore, complex non-equilibrium collisional ionization processes have been calculated for a number of physical processes, including supersonic shocks, conductive interfaces, and turbulent mixing layers between cool clouds and a hot ambient medium \citep{allen08, gnat10, kwak11, wakker12}.  Here, we explore a physical scenario in which SiIV and CIV trace so-called ``transitional temperature" 10$^{5}$ K gas, and are produced predominantly via collisionally-driven cooling from a hot, ionized coronal plasma \citep[e.g.][]{savage09}.  

 The physical conditions that would give rise to a scenario in which, e.g., SiII, SiIII, CII, and CIII are photoionized, but SiIV and CIV are collisionally-ionized, would involve an ionizing spectrum that does not produce sufficient photons with energies $>$ 45 eV. In their study of HVCs exposed to a radiation field due to Galactic ionizing photons and the extragalactic background, \cite{fox05} find that it is difficult to put both CIV and OVI into the same photoionized phase as the low ions. While this may result from CIV and OVI simply arising in lower-density material that is nonetheless still photoionized, it may instead be indicative of collisional processes at work. Figure 8 of \cite{fox05} shows several ionization ``edges" in their assumed spectrum, and a particularly steep drop at 54 eV caused by the HeII edge in hot stars. There are still photons with energies $>$ 54 eV, but the uncertainty in their overall flux is large enough to make the collisional ionization scenario for SiIV and CIV plausible. In our analysis that follows, we ignore photoionization and require the majority of the SiIV and CIV to be collisionally-ionized in a $\sim$10$^{5}$K medium. 

Figure \ref{fig:collisionionfrac} shows predictions of gas column densities and ionization fractions from collisional ionization models both in and out of equilibrium \citep{gnat07, gnat10}. In CIE, as shown with the black curve on Figure \ref{fig:collisionionfrac},  the fraction of C in CIV at these temperatures is $\sim$15\% for the observed range of log N$_{\rm SiIV}$/N$_{\rm CIV}$. This fraction implies a total vertical carbon column density of log N$_{\rm C, total}$sin$b$ $\approx$ 14.9.  Non-equilibrium models seen in brownish-red and green on Figure \ref{fig:collisionionfrac} suggest an even higher total Carbon column density. 

We now turn to the implied total Hydrogen column in the supposed tenuous gas phase in CIE for a range of gas-phase metallicities.  If the gas is metal-enriched at the solar level, this implies log N$_{\rm H}$sin$b$ $= $ 18.4, or $\sim$5\% of the total N$_{\rm H}$sin$b$ implied by pulsar dispersion measure estimates described in the previous Section. If this gas is instead at [M/H] $=$ 0.3, the amount of hydrogen would increase to log N$_{\rm H}$sin$b$ $= $ 18.9, or $\sim$ 16\% of the total N$_{\rm H}$sin$b$ measurement in the Milky Way thick disk. {\emph { Given well-reasoned estimates by \cite{howk06} that the log N$_{\rm H, HIM}$sin$b$ $\approx$ 19.0, the CIE model strongly implies a  [M/H] $>$ 0.3 for the warm and hot components of the Milky Way halo.}}   We further note that the lower f$_{\rm CIV}$ values implied by non-equilibrium models with [M/H] ranging from 0.1 - 1.0 (brownish-red and green curves in Figure \ref{fig:collisionionfrac}) would require N$_{\rm H}$sin$b$ $>$ 19.4, or that $>$ 50\% of the ionized hydrogen column in the halo be in this collisionally ionized 10$^{5}$ K phase. This picture is not physical, and would directly contradict the observational evidence that the WIM and the HIM are likely dominant contributors to the line-of-sight N$_{\rm e}$ \citep{reynolds93,gaensler08, howk06}. 

Previous considerations of collisionally ionized 10$^{5}$K gas in the MW halo have envisioned layers and/or interfaces \citep[e.g.][]{fox09, wakker12}. In these structures,  either conductive interfaces or turbulent mixing layers, predictions of log N$_{\rm SiIV}$/N$_{\rm CIV}$ lie in our observed range, but the total CIV column densities of each layer are closer to $\sim$10$^{12}$ cm$^{-2}$, two orders of magnitude lower than we observe \citep{kwak11}. This inconsistency is alleviated if we demand our line of sight intersects hundreds of such overlapping layers. However, in this scenario, hundreds of layers cooling out of equilibrium would tend to imply values of log N$_{\rm H}$sin$b$ $>$ 19.4 in this 10$^{5}$K ``tenuous" gas phase, just as we described in the previous paragraph.  Furthermore,  hundreds of interfaces or layers between small cool clouds and the HIM are difficult to reconcile with the $\sim$kpc-scale spatial correlations of the SiIV and CIV in the xy plane.

Gas cooling behind supersonic shocks has also been invoked to explain the presence of highly-ionized species such as SiIV and CIV in the circumgalactic medium \citep{dopita96, werk16, mcquinn17}. Under the assumption of a steady, one-dimensional flow, and a given metallicity, several models predict column densities of high and intermediate ions in the post-shock cooling layers. Here we use the results of
\cite{allen08} who produce tables for gas at solar metallicity,  shock speeds that range from 100 $-$ 1000 km s$^{-1}$ and pressure exerted by the transverse magnetic field (Bn$^{-\frac{1}{2}}$) ranging from 10$^{-4}$ $\mu$G cm$^{-3/2}$ - 10 $\mu$G cm$^{-3/2}$. The magnetic pressure is important because it limits the compression through the shock, and thus higher magnetic parameters produce higher gas column densities.  These shock models predict values of log N$_{\rm SiIV}$/N$_{\rm CIV}$ to range widely from -1.0 to 0, mostly dependent upon the shock speed.  However, when we consider the predicted values of the gas column densities at the observed ratio of -0.7, they are closer to a few times 10$^{12}$ cm$^{-2}$ and 10$^{13}$ cm$^{-2}$ for SiIV and CIV, respectively, or more than an order of magnitude lower than we observe. 

We conclude that the collisional ionization case is most plausible if the intermediate ionization gas is in collisional ionization equilibrium, presumably cooling from the hot coronal medium.  In this picture, the 10$^{6}$ K gas and the 10$^{5}$ K gas would have the same metallicity, and the majority of the cooling would be taking place within 3 kpc of the MW disk.   The contribution of the HIM to the value of N$_{\rm e}$ is uncertain, but if we return to the discussion in \S \ref{sec:ionmodel}, we may use log N$_{\rm H, HIM}$sin$b$ $\approx$ 19.0 \citep{howk06}. Using the corresponding constraint on log N$_{\rm C, total}$sin$b$ $\approx$ 14.9 (Figure \ref{fig:collisionionfrac}) in the collisionally ionized medium (CIM), and requiring that N$_{\rm H, CIM}$ $<$ N$_{\rm H, HIM}$,  we find that the hot corona of the Milky Way must have  [M/H] $>$ 0.3 in this physical scenario.

\subsection{Estimate of Areal Mass Infall Rate}
\label{sec:mass}
Given the above constraints on the physical conditions of the CIV-bearing gas, we may estimate areal gas accretion rate for the observed region of sky containing our BHB lines of sight, parameterized as: \begin{equation} \frac{\rm d \it M_{\rm gas}}{\rm d\it t \rm d \it A} \equiv \frac{ M_{\rm gas}v_{\rm infall}}{d_{\rm gas} \pi R_{\rm cloud}^{2}} \left[\rm M_{\odot} ~\rm kpc^{-2}~ \rm yr^{-1}\right]. 
 \end{equation} We determine the total gas mass from column density measurements, using an effective surface area calculation: \begin{equation} M_{\rm gas} = 310 \left( \frac{Z_{\odot}}{Z}\right)\pi R_{\rm cloud}^{2} \left< N_{\rm CIV}\rm sin \it b \right>m_{\rm C} \left( \frac{1}{f_{\rm CIV}}\right) \left[\rm M_{\odot}\right]. \end{equation} Equation 4 assumes the solar abundance of Carbon, 12 $+$ log (C/H), to be 8.43 \citep{asplund09}. Among the quantities relevant to this calculation, we allow for the following values and ranges of values:  (1) the vertical component of the mean CIV column density of the IV gas, $\left< \rm N_{\rm CIV} \rm sin \it b\right>$ $=$ 10$^{14.04}$ cm$^{-2}$;   (2) the fraction of carbon in CIV, 0.09 $<$ f$_{\rm CIV}$ $<$ 0.17;   (3) the gas-phase metallicity, 0.3 $<$ Z/Z$_{\odot}$ $<$ 1.0;  (4) the physical size scale of the absorbing structure R$_{\rm cloud} \approx$ 1 kpc; (5) the vertical infall velocity of the IV component of CIV absorption, v$_{\rm infall}$ $\equiv$ $-v$sin$b$ $\approx$ $45$ km s$^{-1}$; and (6) the distance to the absorbing structure, 1.0 kpc $<$ d$_{\rm gas}$ $<$ 3 kpc.   We note that this areal mass infall rate considers only the material in this intermediate-ionization state gas containing CIV and SiIV, but captures constraints from both photo and collisional ionization within the range of allowed f$_{\rm CIV}$ values and metallicities.   
 
 Following Equations 3 and 4,  the resulting bounds on the areal mass accretion rate in warm, ionized gas at high galactic latitudes are:  \begin{equation} 0.0003 < \frac{\rm d \it M_{\rm gas}}{\rm d\it t \rm d \it A}  < 0.006 \left[  \rm M_{\odot} ~\rm kpc^{-2}~ \rm yr^{-1}\right]. \end{equation} This range of areal mass accretion rate encompasses the mean star formation rate surface density of the Milky Way disk, $\left< \Sigma_{\rm SFR, MW} \right>$  = 0.0033 M$_{\odot}$ yr$^{-1}$ kpc$^{-2}$ \citep{kennicuttandevans12}.  Our implied total mass infall rate of high-latitude WIM,  d$M_{\rm gas}$/d$t$, is 0.02 M$_{\odot}$ yr$^{-1}$ at most and at least 0.001 M$_{\odot}$ yr$^{-1}$. This range is lower than estimates from all-sky studies of the mass infall rates from the ionized components of HVCs and IVCs, excluding the Magellanic System, that find ranges of d$M_{\rm gas}$/d$t$ to be  0.45 $-$ 1.40 M$_{\odot}$ yr$^{-1}$ at d $<$ 15 kpc \citep{lehner11, fox19}. It is also considerably lower than the mass infall rate for the entire MW halo,  $\gtrsim$ 5 M$_{\odot}$ yr$^{-1}$,  \citep{fox14, richter17}. It is unlikely that the small neutral clumps we have observed in this region provide a substantive boost to this mass infall rate \citep[e.g.][]{putman12}.  However, it is possible that material from elsewhere in the halo makes it to the disk via other routes followed by radially mixing within the disk \citep[e.g.][]{sellwood02}.  In contrast, our observations may instead imply that much of the more distant ionized halo material that is moving toward the MW disk never in fact makes it through disk-halo interface to eventually form new stars \citep{joung12}.

\section{Summary and Dicussion}
\label{sec:discussion}

\subsection{Summary of Key Results}
\label{sec:summary}

Our data provide novel, independent constraints on the coherence scales of IV gas in low and intermediate ionization states at the disk-halo interface. We find that CIV and SiIV are strongly correlated over the full range of transverse separations probed by our sightlines ($\sim$40$^{\circ}$ or 2 kpc at z $=$ 3 kpc) and with N$_{\rm HI}$ from 21-cm emission. In contrast,  the low ionization state material traced by CaII and FeII neither exhibits a correlation with N$_{\rm HI}$ nor transverse separation (Figures \ref{fig:nionnhi} and \ref{fig:deltacol}). This readily apparent difference in ionic behavior indicates the existence of large, coherent ionized structures $\gtrsim$ 1 kpc  in size, containing cool, neutral cloudlets  or clumps with spatial scales $\lesssim$ 10 pc (10.8$\arcmin$  at z $=$ 3 kpc). We explore the implications of this result in the next Section, \S \ref{sec:sizes}.  Finally, we find that the implied areal mass infall rate of the high latitude WIM encompasses the present-day mean star formation rate surface density of the Milky Way disk, although is slightly lower that previous estimates from all-sky data. 
 
The SiIV/CIV ratio of the IV gas along our sightlines, -0.69$\pm$0.04, shows remarkably little variation over $\sim$40
degrees. We argue that this column density ratio places two
mutually-exclusive constraints on the physical conditions of the extraplanar gas
under two distinct sets of assumptions: photoionization and collisional ionization. In the photoionization case:  \begin{enumerate}
\item{We assume that SiIV and CIV are produced by photoionization from a combination of starlight
from the Milky Way disk and the extragalactic UV background at z$=$0
\citep{fox05}. }
\item{We adopt constraints from in-depth work that constrains the cool,
low-ionization state IV gas to have a metallicity of approximately
solar value, requiring that cool and warm gas are well-mixed in terms
of their metal content \citep{wakker08}.}
\item{Finally, we assume that the total Hydrogen column associated with
the SiIV and CIV is given by the total electron column measured from pulsar dispersion measure experiments at high galactic latitude, with appropriate corrections for He, the HIM, and the neutral gas fraction.  }
\end{enumerate}
Proceeding in this framework of a photoionized WIM, we find that log N$_{\rm SiIV}$/N$_{\rm CIV}$ $\approx$ -0.7 requires that
Si be depleted relative to C by 0.25 $-$ 0.35 dex compared to the solar value
of Si/C. Under these assumptions, this paper would present  observational evidence for a dusty WIM at the Milky Way disk-halo interface. We explore the implications of this result in \S \ref{sec:dust}.

The other constraint on the physical state of the material relies only on the production of SiIV and CIV by collisional ionization in a 10$^{5.2}$ K gas phase cooling out of a hotter, 10$^{6}$ K ambient medium with a density $<$ 8 $\times$ 10$^{-4}$ cm$^{-3}$ \citep[e.g.][]{bregman07}. When we consider the implied total Hydrogen content of this transient, cooling phase, we find that there is significant tension if we allow halo [M/H] $<$ 0.3. Thus, in the collisional ionization framework, our constraint on the total carbon content implies a Milky Way corona that is enriched at a level of  greater than a third solar. This constraint would represent the first observational constraint on the metallicity of the hot corona of the Milky Way. 

Additional observations are required to determine which of these two models of the physical state of the IV gas is accurate. For example, access to doubly ionized species such as SiIII, SIII, AlIII, and FeIII would better constrain the photoionization models, and possibly provide further evidence of dust depletion in the ionized extraplanar material (e.g. Howk \& Savage 1999). The detection of more highly ionized species such as OVI, OVII, or OVIII in the IV gas would be crucial for a complete understanding of the nature of collisionally ionized material in the halo \citep{faerman17}. Finally, our results underscore the importance of ionization modeling that considers multiple gas phases under different ionization conditions, with the additional complicating consideration of significant depletion onto dust in ionized extraplanar material. Thus, to convincingly model the physical state of diffuse gas requires observers to study as many ionic species as possible along any given sightline.

\subsection{Cloud Length Scales and Physical Models}
 \label{sec:sizes}

We find that intermediate-velocity absorption from CIV and SiIV likely traces a coherent ionized structure at the disk-halo interface spanning at least $\sim$1 kpc across. The much larger scatter in the low ions (e.g. CaII) on the same spatial scales is consistent with a picture in which cool clumps of gas are considerably more compact ($<$ 10 pc in length scale). All of this gas most likely lies within 3.4 kpc of the MW disk, and may contain enough total mass to maintain the current star formation rate of the Milky Way. 

 \subsubsection{Comparison with Complimentary Observational Analyses}
 
Small, compact neutral structures are observed directly via 21-cm emission in the Milky Way halo. In particular, as one improves the spatial resolution with which the neutral gas in the Milky Way halo is observed, small, pc-scale structures are detected with increasing frequency \citep{lockman02, saul12}. Our work finds both ionized and neutral phases are detected along all seven sightlines, moving at similar velocities, indicating that these small, cool clumps are likely contained within a larger-scale highly-ionized structure moving toward the Milky Way disk at a velocity of 30$-$50 km s$^{-1}$.  Previous work has highlighted the discrepant covering fractions of the low and intermediate ionization state material in the Milky Way halo \citep{lehner12,richter17} as evidence supporting the multiphase structure of the gas. Our evidence for this multiphase structure of the inner Milky Way halo is both complimentary and more direct.  In addition, our physical interpretation of the phase structure is consistent with that of \cite{fox04},  who find the high-density neutral HVCs to be surrounded by much larger envelopes of diffuse, ionized material in absorption-line experiments. 
 
The physical scenario in which both intermediate and low ionization state gas is photoionized by a combination of the EUVB and star formation in the MW disk \citep{fox05} therefore requires that at least two distinct gas phases characterized by different densities (or length scales) be considered. One straightforward means of modeling multiphase, photo-ionized gas is to have the multiple phases arise from a smooth gas density gradient in the absorbing material. \cite{stern16} find that an outwardly declining density gradient of a factor of 100 per kpc can successfully reproduce the multiphase ion column densities from MgII to OVI in  L* halos at z$\sim$0.2. This density gradient implies length scales of MgII-traced material to be $\sim$50 pc, while SiIV-traced material has a characteristic length scale of 3.9 kpc, a factor of $\sim$80 difference. We find a  factor of $\approx$ 100 difference in the length scales of CaII-traced (10 pc) and CIV (SiIV)-traced gas (1 kpc), suggesting a similarly steep gas density gradient if the multiphase gas in the Milky Way disk-halo interface is predominantly photo-ionized. 

Recently, \cite{zheng18} conducted a careful, statistical analysis of the SiIV absorption along 132 sightlines that pierce the Milky Way disk-halo interface and found evidence for two components contributing to the absorption: (1) a flat-slab component similar to that of \cite{savage09} and (2) a global component analogous to that traced by ionized metals observed along extragalactic lines of sight toward background QSOs within projected distances of 10 - 150 kpc from  L* galaxies \citep[e.g.][]{werk13}. It is primarily the lack of a correlation between Galactic latitude and N$_{\rm SiIV}$ that allows them to disfavor a flat-slab model alone, without some global contribution. Explicitly, the flat-slab model tends to predict larger logN at lower $b$ because path-lengths at lower latitudes are longer; such a trend is distinctly absent from the data. Additional model parameters such as patchiness tend to confound the interpretations, however. 

The two-component model of \cite{zheng18} predicts that the disk-halo component would contribute log N$_{\rm SiIV}$ $=$ 12.1 at high Galactic latitudes, whereas our data imply log N$_{\rm SiIV}$ $=$ 13.35 toward the NGP within 3.4 kpc. Our results, then, appear to be at odds with the analysis of \cite{zheng18}.  However, we note that relative column density contributions from the disk-halo interface and global components are rather poorly constrained in the two-component model, especially when one allows for patchiness or clumpiness.  Our data consist of only 7 sightlines toward high latitude BHB stars, and therefore cannot rule out the existence of a global component as put forth by \cite{zheng18}, especially if the column density of nearby gas dominates over the global component. However, our observations do seem to imply that the relative contribution of the global component to ionized gas column densities in the MW halo are probably lower than that reported by  \cite{zheng18}.  Indeed, because \cite{zheng18} include only QSO sightlines in their analysis, they likely underestimate the contribution from nearby gas to the observed column densities. Their model could be improved  by adding in nearby stellar sightlines, such as those analyzed in this study,  to achieve better control of the parameter fitting.  

\subsubsection{Comparison with Simulations}

 Observed limits on length-scale, prevalence, velocity and density of cool, extraplanar clouds with z $<$ 3.4 kpc can ultimately constrain detailed feedback models that seek to explain the origin and influence of galaxy-scale winds. For example, small clumps $<$ 10 pc in size may affect the relevant cooling processes and thus evolution of material at the disk-halo interface. Recent progress in hydrodynamical modeling has led to increasingly sophisticated treatments of diffuse halo gas that move well beyond the Galactic fountain as it was originally envisioned, in which spontaneous radiative instabilities cause clouds to cool and  fall back down onto the disk ballistically \citep{shapiro76, bregman80}.  Such idealized, high resolution simulations of Galactic fountain processes, guided by a range of observational constraints, are well-poised to address the uncertain physics at the disk-halo interface \citep[e.g.][]{marinacci11, fraternali17,armillotta17, schneider18, fielding18, gronnow18, gronke18, kim18, sormani19, melso19}. These models include a wide range of complex physics that impact the gas evolution to varying degrees:  magnetic fields, halo rotation, conductive interfaces, and stratified turbulent layers. 
 
 Here, we highlight a few case studies of disk-halo interface or fountain gas to show the promise of high-resolution idealized simulations in untangling the complex gas physics at work in this regime. In the ``Three-Phase ISM in Galaxies Resolving Evolution with Star fomation and SN feedback" (TIGRESS) simulation suite, \cite{kim18} present a comprehensive, self-consistent model of the multiphase star-forming ISM in which a fountain flow arises naturally in warm-ionized material energized by local SNe. The TIGRESS simulation allows for a phase-by-phase analysis of gas flows in a 3D box 1 kpc $\times$ 1 kpc in the horizontal direction and $-$4.5 kpc $<$ z $<$ 4.5 kpc in the vertical direction, centered on a region of a galaxy disk with parameters similar to those in the solar neighborhood. The TIGRESS set-up is therefore directly analogous to our MW BHB experiment.
 
 Most saliently, the self-consistent treatment of star formation, FUV emission from massive stars, and SNe in TIGRESS generates outflows in a hot phase (T $>$ 5 $\times$ 10$^{5}$ K), and periods of upwelling and subsequent fallback (a fountain flow) in the cool, warm, and ionized phases with T $<$ 5 $\times$ 10$^{5}$ K. The temporal variation in the SFRs leads to $\sim$ 50 Myr periods in which the warm and ionized gas phases are dominated by successive periods of outflow and inflow. Thus, in this rendering of the Galactic fountain, our observations would indicate that the WIM above the solar neighborhood is currently experiencing a temporary inflow period. Furthermore, the gas inflow velocities in the simulations of $\sim$60 km s$^{-1}$ are similar to our values of $v$sin$b$ which range between 40$-$ 50 km/s for the warm ionized gas traced by SiIV and CIV. 
 
 Additionally, TIGRESS provides a natural explanation for why we do not observe substantial outflowing material in conjunction with inflowing material in our area. First,  $>$ 70\% of the outflowing gas by mass is in a hot phase inaccessible by our COS observations. Second, although inflows and outflows are contemporaneous in the simulations in every phase, warm gas, when outflowing, is pressure confined by the hot gas into small cloudlets, which would have a comparably small covering fraction. Once an episode of star formation ceases, the warm ionized material falls back and expands to fill nearly an order of magnitude more volume than when it is in the outflowing period. 
 
 Finally, we note the areal mass infall rate in the warm ionized fountain flows in the TIGRESS simulations range between 0.001 $-$ 0.01 M$_{\odot}$ yr$^{-1}$ kpc$^{-2}$ between 1 and 3 kpc above the disk, which is similar to the time-averaged star formation rate of their simulated disk of $\sim$0.006 M$_{\odot}$ yr$^{-1}$ kpc$^{-2}$. Our areal mass infall rate estimated in Section \ref{sec:mass} is similar, 0.0003 $-$ 0.006 M$_{\odot}$ yr$^{-1}$ kpc$^{-2}$ within 3 kpc. We note that several facets of the simulations impair such direct comparisons with observations. TIGRESS does not include photoionization, so it is not clear how the gas column densities we measure (e.g. N$_{\rm CIV}$) translate to a warm, ionized phase in the model. In addition, TIGRESS does not include a global geometry for the simulated galaxy, which may matter for any larger-scale gas flows, especially those that may originate in the center of the galaxy where star formation is more concentrated.  
 
 Other high-resolution idealized simulations inform cloud survival times and mass loading rates of the gas flows in the Galactic fountain. \cite{armillotta17} find from extremely high-resolution two-dimensional simulations of the evolution of cool clouds passing through an ambient hot medium that the characteristic length scale of cool gas clumps will influence their survival times in a halo environment, such that cool clouds $>$ 250 pc in size are able to retain their cool gas mass (although they are subject to fragmentation due to thermal instabilities) over $>$ 250 Myr timescales. However, the recent analysis of three-dimensional simulations of cold cloud evolution by \citet{gronke18} concluded that the survival time of these structures is in fact governed by the cooling time of ``mixed'', intermediate-temperature material surrounding the cloud, rather than by the formal cloud-crushing time.  This enables the survival of much smaller cold structures with lengths $\gtrsim 2$ pc, and could potentially explain the existence of such small structures consistent with our observations. In the \citet{fielding18} study of clustered SNe-driven winds in idealized three-dimensional hydrodynamical simulations of sections of galaxy disks,  the ability of superbubbles to vent into the halo is regulated in part by the disk scale height.  Once breakout occurs, the wind mass loading is determined by the mixing and cooling of material stripped off of entrained cold clouds \citep{fielding18}. Finally, the CGOLS suite of high-resolution isolated galaxy models likewise demonstrates that interactions between fast-moving outflow material and small, cool, dense clouds can lead to increased mass-loading of the hot phase, which has dramatic implications for the impact of feedback in galactic halos \citep{schneider18}.  
 
 We note that length scales $<$ 100 pc are not currently resolvable in state-of-the-art cosmological hydrodynamical models that aim to probe diffuse halo gas. Such simulations now reach spatial resolutions in the CGM on the order of hundreds of parsecs \citep{peeples18, hummels18, vandevoort19, suresh19}, and find that even a factor of $\sim$10 increase in spatial resolution can produce dramatic effects on the physical processes and phase structure of the gas. In these simulations,  there is a substantial neutral hydrogen abundance in clouds with sizes $<$ 1 kpc
 (and such high-density clouds are predicted to occur on even smaller scales as resolution is increased yet further).  Our results support the conclusions of these studies, namely that higher resolution simulations in a cosmological context are required to trace the relevant physical processes and phase structure of the CGM and disk-halo interface.  Recent progress, however, indicates that a full understanding of the Galactic fountain, including its timescales, extent, and mass and energy loading factors, is on the horizon.

\subsection{A Dusty WIM}
 \label{sec:dust}

Our photoionization modelling of extraplanar SiIV and CIV column densities suggests the presence of Si-bearing dust in the high-latitude WIM. Silicon depletions on the order of a few tenths of a dex are consistent with levels of depletion in DLAs of $\delta_{\rm Si}$ = $-$0.27$\pm$0.16 dex (measured with Si/Zn) assuming Carbon is not depleted itself \citep{vladida11}.  Si/C ratios of $\sim$-0.7 appear to occur in the neutral Milky Way ISM at a value of F* of 0.3 $-$ 0.5, where F* is the line of sight depletion strength factor \citep{jenkins09} that increases with increasing depletion. Notably, the 1$\sigma$ sightline-to-sightline variation in Si/C we observe is only $\pm$ 0.04 dex. Under the assumption of photoionization, this constant ratio would indicate remarkably uniform dust coverage over the kpc-sized spatial scales of the material traced by SiIV and CIV. 

We note that depletion onto dust from ionized material is not entirely unexpected.  \cite{howk99}  discovered the first evidence for dust in the diffuse ionized gas of the Galactic ISM from modeling the observed ratio of AlIII/SIII, which traces the ratio of a depleted element (Aluminum) to a non-depleted element (Sulfur) in the ISM . Their result, like this one, is contingent upon SIII and AlIII not having significant contributions from collisional ionization. Further confirmation of dust in an ionized phase came from \cite{lagache99, lagache2000} who first detected far-IR emission from high-latitude dust using COBE and HI data, and refined this measurement with data from the Wisconsin H$\alpha$ Mapper (WHAM). Additionally, \cite{dobler09} find evidence for a spinning dust component of the Milky Way WIM by analyzing the H$\alpha$-correlated emission spectrum of the five-year WMAP data. Finally, a study by \cite{gringel00} found an absorption component of molecular Hydrogen in an ORFEUS-Echelle spectrum at $-62$ km s$^{-1}$, which they associated with the Milky Way halo, a sign that there are grains present upon which H2 can form. Both absorption and emission studies support a picture in which dust temperature and emissivity is similar between neutral and ionized gas, and that the processing of dust grains in the WIM is similar to that in the HI gas. 

Dust may persist up to several Mpc from the disks of galaxies, as traced by a reddening coefficient similar to that of interstellar dust \citep{menard10, peek15}. Furthermore, a substantial fraction of the metals generated within the galaxy may be lurking in a circumgalactic dust phase \citep{peeples14}. The coexistence of dust and highly ionized gas in galactic coronae has far-reaching implications for self-shielding, magnetic fields, molecule formation, gas cooling, and grain destruction in rarefied environments. Future studies of gas physics in the disk-halo interface and the CGM should incorporate the physics of dust, and allow for non-solar ratios of the elements to accurately interpret observations with hydrodynamical simulations.

\section{Acknowledgements}
   
 Support for this work was provided by NASA through program GO-14140. JKW and HVB acknowledge additional support from a 2018 Sloan Foundation Fellowship, and from the Research Royalty Fund Grant 65-5743 at the University of Washington. AD is supported by a Royal Society University Research Fellowship. We thank B. Winkel for help with the processing of the EBHIS data and Chris Howk, Mary Putman, Josh Peek, Karin Sandstrom, Dusan Keres, Drummond Fielding, Evan Schneider, Chang-Goo Kim, Max Gronke, and Todd Tripp for helpful input and discussions regarding the results from this study. YZ acknowledges support from the Miller Institute for Basic Research in Science.  AD and JKW thank Risa Wechsler and Charlie Conroy for organizing a speed collaboration session at the Mayacamas Ranch Conference in 2015, where the idea for this project originally took shape. Finally, we thank the anonymous referee who provided valuable, in-depth comments that greatly improved the manuscript. 
 
 The optical data presented herein were obtained at the W. M. Keck Observatory, which is operated as a scientific partnership among the California Institute of Technology, the University of California and the National Aeronautics and Space Administration. The Observatory was made possible by the generous financial support of the W. M. Keck Foundation.  The authors wish to recognize and acknowledge the very significant cultural role and reverence that the summit of Maunakea has always had within the indigenous Hawaiian community.  We are most fortunate to have the opportunity to conduct observations from this mountain. \\
  {{\it Facilities:} \facility{HST: COS} \facility{Keck: HIRES}} \\    
  {{\it Software:} CALCOS (https://github.com/spacetelescope/calcos/); MPFIT1 (http://cow.physics.wisc.edu/$\sim$craigm/idl/fitting.html); CLOUDY (https://www.nublado.org/) }

\bibliographystyle{apj}
\bibliography{bhbrefs}

\end{document}